\documentclass[floatfix,%
 aip,
 amsmath,amssymb,
 reprint,%
]{revtex4-1}
\usepackage{graphicx}
\usepackage{amsmath}
\usepackage{hyperref}
\usepackage{booktabs} 
\usepackage{placeins}

\usepackage{graphicx}
\usepackage{dcolumn}
\usepackage{bm}
\usepackage{xcolor}
\usepackage[dvipsnames]{xcolor}

\usepackage[utf8]{inputenc}
\usepackage[T1]{fontenc}
\usepackage{mathptmx}
\usepackage{etoolbox}
\usepackage{subcaption}

\usepackage{newunicodechar}

\begin{document}

\title{Bayesian Optimization of Laser-Wakefield Acceleration via Spectral Pulse Shaping}
\author{B. Z. Djordjevi\'{c}}
\affiliation{ Lawrence Livermore National Laboratory, Livermore, California 94550, USA}%
\email{djordjevic3@llnl.gov}
\author{C. Benedetti}%
\affiliation{ Lawrence Berkeley National Laboratory, Berkeley, California 94720, USA}%
\author{A. D. McNaughton}%
\affiliation{ Lawrence Livermore National Laboratory, Livermore, California 94550, USA}%
\affiliation{ Portland State University, Portland, Oregon, 97201, USA}%
\author{R. Lehe}%
\affiliation{ Lawrence Berkeley National Laboratory, Berkeley, California 94720, USA}%
\author{H.-E. Tsai}%
\affiliation{ Lawrence Berkeley National Laboratory, Berkeley, California 94720, USA}%
\author{S. C. Wilks}%
\affiliation{ Lawrence Livermore National Laboratory, Livermore, California 94550, USA}%
\author{B. A. Reagan}%
\affiliation{ Lawrence Livermore National Laboratory, Livermore, California 94550, USA}%
\affiliation{ Department of Electrical and Computer Engineering, Colorado State University, Fort Collins, CO.80523, USA}%
\author{G. J. Williams}%
\affiliation{ Lawrence Livermore National Laboratory, Livermore, California 94550, USA}%
\author{J. van Tilborg}%
\affiliation{ Lawrence Berkeley National Laboratory, Berkeley, California 94720, USA}%
\author{C. B. Schroeder}%
\affiliation{ Lawrence Berkeley National Laboratory, Berkeley, California 94720, USA}%
\affiliation{ Nuclear Engineering Department, University of California, Berkeley, California 94720, USA}%


\begin{abstract}
In this paper, we investigate the effect of spectral pulse shaping of the laser driver on the performance of channel-guided, laser-plasma accelerators. The study was carried out with the assistance of Bayesian optimization using particle-in-cell simulations. We used a realistic plasma profile based on a novel optical-field-ionized channel technique with ionization injection and low, on-axis plasma densities to maximize the energy gain of the electron bunch trailing the laser. Spectral shaping allows us to modify the temporal profile of the laser driver while keeping the laser energy constant, affecting the acceleration and injection processes. In addition we consider how modifying the plasma channel parameters may affect the target outputs. Given the complexity and breadth of the parameter space in question, we used numerical optimization to identify high performers. In particular, we found laser profiles with additional spectral content that, when used with optimal plasma channel parameters, result in charge content an order of magnitude higher than the baseline Gaussian case while also increasing the mean energy of the electron bunch.
\end{abstract}


\maketitle

\section{Introduction}
\label{intro}

Laser wakefield acceleration (LWFA) has emerged as a leading platform for generating compact, high-gradient particle accelerators\cite{akhiezer1956,tajima1979,esarey2009}. In LWFA, an intense laser pulse propagating through a plasma drives a wake capable of accelerating charged particles to gigaelectronvolt (GeV) energies over tens of centimeters.\cite{picksley2024,rockafellow2025} This technique promises revolutionary applications in particle physics,\cite{schroeder2010,schroeder2016} medical imaging,\cite{cole2015}, light sources,\cite{albert2016,barber2025} and high-energy-density science.\cite{albert2024} A recent development is the use of LWFA generated electrons as a source for muons to be used as an advanced diagnostic.\cite{terzani2025,ludwig2025}

Over the past two decades, LWFA has seen rapid progress in experimental performance\cite{geddes2004,leemans2006,gonsalves2019,miao2022} modeling capability,\cite{lehe2015,vay2016,shapoval2024} and the underlying theory.\cite{lu2006,tzoufras2008,esarey2009} Key to these advances has been the development of tailored plasma channels that guide the laser pulse and sustain an accelerating and focusing structure.\cite{durfee1993,geddes2004,gonsalves2019} Recent innovations such as hydrodynamic optical-field-ionized (HOFI) plasma channels have enabled unprecedented control over the laser-plasma interaction, reducing diffraction and permitting longer acceleration lengths~\cite{shalloo2020,miao2020}. These advances have recently culminated in the demonstration of electron energies of 10~GeV in a single LWFA stage ~\cite{picksley2024,rockafellow2025,aniculaesei2023}, establishing a new benchmark in the field.

Despite these achievements, improving beam quality and maximizing injected charge remain major challenges. The complex, high-dimensional parameter space associated with LWFA, e.g., laser pulse shape, plasma profile, injection dynamics, etc., makes systematic optimization difficult using traditional trial-and-error approaches or grid scans. As a result, numerical optimization techniques, particularly those based on machine learning, have gained traction. Controlling the spectral phase of the laser pulse for short-pulse laser plasma interactions in particular, whether to maximize the compression or tune the pulse shape, is challenging to do manually and has benefited from various numerical optimization techniques already.\cite{streeter2018,dann2019,shalloo2020,mariscal2024}

Bayesian optimization (BO) with Gaussian process (GP) regression offers a powerful framework for guiding simulations or experiments toward optimal configurations with minimal evaluations.\cite{frazier2018} The algorithm iteratively: (1) fits a Gaussian process to all available observations, (2) uses this GP to define an acquisition function that quantifies the expected utility of sampling at each point, and (3) evaluates the objective at the point maximizing this acquisition function. By building surrogate models of the objective landscape, GP-based optimization can identify trends, extrapolate performance, and efficiently navigate sparse parameter spaces. Unlike other techniques such as neural networks or random forests, GPs come with built-in uncertainty quantification, where the posterior variance tells us how uncertain the model is at any point. This approach has been successfully applied in plasma-based acceleration to optimize laser phase, experimental spectral shaping,\cite{ziegler2021, loughran2023} simulations,\cite{valenta2025} injection mechanisms,\cite{backhouse2025} plasma density profiles,\cite{jalas2021} and multi-objective optimization of LWFAs.\cite{irshad2023}

In this work, we apply Bayesian optimization to the problem of tailoring both the laser driver profile and plasma channel to maximize the quality of the accelerated bunch. Our focus is on enhancing the bunch charge and energy while minimizing energy spread using a spectrally-shaped laser pulse coupled to a tapered HOFI plasma channel. We leverage Gaussian process surrogates to explore the parameter space and identify configurations that outperform conventional laser drivers. The results demonstrate the promise of data-driven optimization for advancing the next generation of compact, tunable particle accelerators.

In this manuscript, we discuss the topic as follows. In Sec. \ref{part1} we discuss the basics of spectral shaping of a short-pulse laser and its effects on wakefield acceleration. In Sec. \ref{part2} we discuss the basics of Bayesian optimization with Gaussian processes. In Sec. \ref{setup} we discuss the setup and numerics of our simulations. In Sec. \ref{part5} we present a baseline Gaussian driver example as a reference, show the effect of spectral shaping on LWFA, apply BO to just spectral shaping, and finally apply BO to spectral shaping and channel properties. In Sec. \ref{partexamples} we examine the properties of some particularly high-performing case examples discovered during BO. And finally in \ref{conc} we discuss our conclusions and potential future directions.

\section{Background to Spectral Shaping}
\label{part1}

Femtosecond lasers used in LWFA rely on the precise manipulation of ultrashort optical pulses to drive relativistic plasma waves. These pulses typically span tens of femtoseconds in duration to match the characteristic wavelength of the plasma and possess broadband frequency content as a result of the Fourier limit, i.e., $\Delta t\cdot \Delta f\approx 0.441$ for a Gaussian. The temporal shape of the pulse is intrinsically connected to its spectral amplitude and phase. Although conventional Gaussian pulses are often used in LWFA studies, realistic pulses are never truly Gaussian, either longitudinally or transversely, and recent studies have demonstrated that tailoring the spectral phase and amplitude, which we have termed spectral shaping, can yield significant performance improvements in particle acceleration~\cite{shalloo2020, ziegler2021}.

\subsection{Spectral Shaping of Short Pulse Lasers}

\begin{figure*}[t]
        \centering
\includegraphics[width=0.99\textwidth]{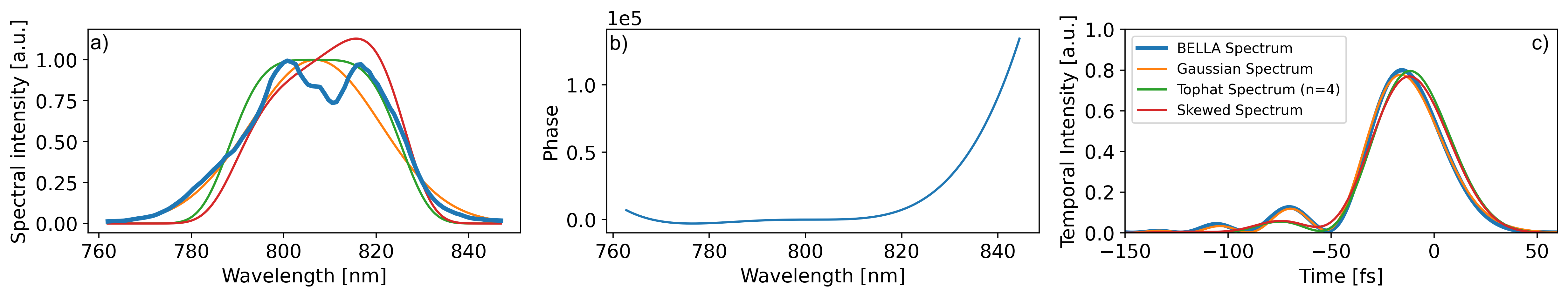}
\caption{\label{fig:spectral} Depiction of the process of spectral shaping. In (a) we can use the physical BELLA spectrum (blue), convolve it with b) a polynomial based on the Taylor expansion of the phase, where we chose the coefficients of the 2nd, 3rd, and 4th order terms, and take the inverse Fourier transform to derive a new, spectrally shaped temporal profile in (c). Keeping the same polynomial phase but using idealized spectra of a Gaussian (orange) or super-Gaussian (green), we can see some variation to the final temporal profile. We also consider a skewed spectrum (red), which can be seen in experiments.}
\end{figure*}

In the frequency domain, the electric field of a femtosecond laser pulse can be expressed as:
$$
\mathcal{E}(\omega) = A(\omega)\, e^{i\phi(\omega)},
$$
where \( A(\omega) \) is the spectral amplitude and \( \phi(\omega) \) is the spectral phase.\cite{trebino2000} The corresponding time-domain electric field \( E(t) \) is obtained via the inverse Fourier transform, that is,
\begin{equation}
\mathcal{E}(t) = \frac{1}{2\pi}\int_{-\infty}^{\infty}\mathcal{E}(\omega)e^{-i\omega t}d\omega = \frac{1}{2\pi}\int_{-\infty}^{\infty}A(\omega)e^{i\phi(\omega)}e^{-i\omega t} d\omega.
\label{iff}
\end{equation}
The Fourier transform and its inverse are performed numerically using a centered inverse fast Fourier transform. We can take a Taylor expansion of the spectral phase to get
$$  \phi(\omega) = \phi(\omega_0) + \phi'(\omega_0)\Delta\omega+\frac{1}{2!}\phi''(\omega_0)\Delta\omega^2 $$
$$
+ \frac{1}{3!}\phi'''(\omega_0)\Delta\omega^3 + \frac{1}{4!}\phi^{(4)}(\omega_0)\Delta\omega^4...,
$$
where $\omega_0$ is the central frequency, somewhat arbitrarily chosen but typically the weighted center of the spectrum, $\phi'$ is the group delay, $\phi''$ is the group delay dispersion (GDD), $\phi'''$ is the third order dispersion (TOD), $\phi^{(4)}$ is the fourth order dispersion (FOD), and $\Delta\omega=\omega-\omega_0$. The Taylor expansion is strictly valid only near $\omega_0$, so for highly structured or broadband spectra, the expansion may be insufficient and a direct manipulation of $\phi(\omega)$ across the entire spectrum may be necessary. The experimental spectrum, seen in Fig. \ref{fig:spectral}.a) (blue), is typically measured with respect to the laser wavelength $\lambda$, which we interpolate into a linear frequency grid through $\omega=2\pi c/\lambda $, and we convolve this with a polynomial phase $\phi(\omega)$, as seen in Fig. \ref{fig:spectral}.b), although the phase can be arbitrarily shaped. 
A Hanning window is used to mitigate the effects of spectral leakage, given that we have a finite spectrum. Once we have $\mathcal{E}(\omega)$, we use Eq. \ref{iff} to derive the modulated temporal profile $\mathcal{E}(t)$, where an example is provided in Fig. \ref{fig:spectral}.c). The intensity $I(t)=|\mathcal{E}(t)|^2$ is scaled to conserve pulse energy, i.e., $\iiint I(t,z,r) r \ dtdzdr = \text{E}_\text{pulse}$. Each laser has its own spectrum, and spectra can vary from shot to shot. Just to illustrate this, we have taken the same approach for an idealized Gaussian spectrum (orange), a super-Gaussian spectrum (green), and a skewed spectrum (red) in Fig. \ref{fig:spectral}. The pure Gaussian spectrum surprisingly has little effect, whereas the super-Gaussian spectrum has a noticeable effect on the prepulse tail. The skewed spectrum is similar to the super-Gaussian in its effects. The caveat here is that one should be careful about generalizing intuition too aggressively when performing spectral shaping.

By controlling \( \phi(\omega) \), one can generate a wide variety of temporal profiles, chirped pulses, asymmetric shapes, double peaks, pulses with extended pedestals, etc. This is regularly done in experiments using a device known as an acousto-optic programmable dispersive filter (AOPDF) or DAZZLER\texttrademark, which is typically used to maximize compression and thereby the intensity. Theoretically, this is done without altering the total energy but, in practice, the spectrum can be truncated and the energy reduced. In this work we will only manipulate the 2nd, 3rd, and 4th order spectral coefficients since this is what is typically done in experiments, but technically we can go to arbitrary order of spectral phase. 

\subsection{Spectral Shaping of Wakefield}

A key application of spectral shaping in LWFA is to manipulate the evolution of the laser pulse during its propagation through plasma. Self-phase modulation, self-focusing, and group velocity dispersion all reshape the pulse in nonlinear ways that can lead to undesirable effects such as premature depletion, pulse breakup, or poor phase matching with the wakefield. Shaping the initial spectral phase can compensate for these effects, preserving the pulse integrity over longer distances. For example, positively chirped pulses may undergo self-compression in a plasma and reach peak intensity only after propagating a distance, extending the effective acceleration length.\cite{kim2017,pathak2018,grigoriadis2023}

We can examine these effects with some basic theory to build intuition. The temporal intensity profile of the laser pulse directly affects the dynamics of plasma wave excitation and electron injection. The ponderomotive force $ F_p \propto \nabla |\mathcal{E}(t)|^2 $ governs the wakefield amplitude and the plasma response can affect the intensity profile. Starting in the quasi-linear limit $(a_0^2 \ll 1)$, the on-axis wake potential $\phi(\zeta,z)$ obeys the following relation, 
$$ \left(\partial_\zeta^2 + k_p^2\right) \phi(\zeta,z)=S(\zeta,z), $$
where $S(\zeta,z)\propto \partial_\zeta^2|a(\zeta,z)|^2$ is the source term, $\zeta = ct -z$, and $k_p =\omega_p/c$. Taking the Green's function solution $\sin[k_p(\zeta - \zeta')]$,\cite{esarey2009} we can express the wake potential via the causal convolution
\begin{equation}
\phi(\zeta,z) =\int_{-\infty}^\zeta  \sin[k_p(\zeta - \zeta')] S(\zeta',z) d\zeta'.   
\end{equation}
To leading order, the source term is proportional to the ponderomotive force, i.e.,
$F_p(\zeta,z) \propto \frac{\partial}{\partial \zeta} |a(\zeta,z)|^2.$
We let the complex laser envelope at $z=0$ be specified by amplitude $a(\omega)$ and spectral phase $\phi(\omega)$. After propagating a distance $z$ in a plasma with dispersion $k(\omega)$, the slowly-varying envelope in the co-moving frame is
$   a(\zeta,z) = \mathfrak{F}^{-1} \left[ a(\omega)e^{i\phi(\omega)}e^{ik(\omega)z} \right], $ where $ \mathfrak{F}^{-1}$ denotes the inverse Fourier transform with respect to $\omega$ onto $\zeta$. The general source entering the Green's function integral is then
\begin{equation}
    S(\zeta',z) \propto \frac{\partial^2}{\partial\zeta^2} \left| \mathfrak{F}^{-1} \left[ a(\omega)e^{i\phi(\omega)}e^{ik(\omega)z} \right] \right|^2.
\end{equation}
We can analytically derive the effect of spectral shaping when considering a Gaussian spectral amplitude centered at $\omega_0$ with a bandwidth of $\sigma_\omega$, i.e., $a_0 \exp(-\omega^2/2\sigma_\omega^2)$ and quadratic spectral phase (linear chirp) $\phi(\omega)=\frac{1}{2}\phi''(\omega-\omega_0)^2$. We can also expand the dispersion as $k(\omega) \approx k_0(\omega_0) + k'(\omega_0)(\omega-\omega_0) + \frac{1}{2}k''(\omega_0)(\omega-\omega_0)^2$, where $k'(\omega_0)=k_1=v_g^{-1} = \frac{dk}{d\omega}|_{\omega_0}$ is the inverse group velocity and $k''(\omega_0)=\beta_2=\frac{d^2k}{d\omega^2}|_{\omega_0}$ is the group velocity dispersion coefficient.\cite{agrawal2013} Up to the carrier phase we can write the inverse Fourier transformed pulse as 
$$
a(\zeta,z) = \int \frac{d\omega}{2\pi} \exp\left[-\frac{\omega^2}{2\sigma_\omega^2} + i \frac{\phi''+\beta_2 z}{2}\omega^2 - i(\zeta-k_1z)\omega \right].
$$
Performing the Gaussian integral shift, $\zeta \rightarrow \zeta - k_1 z$, with $\tau_0=1/\sigma_\omega$, we have
$$
C_\text{eff}(z) = \frac{\phi''+\beta_2z}{\tau_0^2}, \ \ \ \tau^2(z) = \tau_0^2[1+C_\text{eff}^2(z)],
$$
and the propagated laser envelope for a chirped Gaussian with dispersion included,
\begin{equation}
    a(\zeta,z) = a_0 \frac{\tau_0}{\sqrt{1+C_\text{eff}^2}} \exp\left[ -\frac{\zeta^2}{2\tau^2(z)}\right]\exp\left[ i \frac{C_\text{eff}(z)}{2\tau^2(z)}\zeta^2\right].
\end{equation}
We can then write the acceleration field as 
\begin{equation}
    \frac{\mathcal{E}_\text{max}(z)}{\mathcal{E}_0} = \frac{\sqrt{\pi}}{4} a_0^2 k_p \tau(z) \exp\left[ - \frac{k_p^2\tau^2(z)}{4}\right],
\end{equation}
where $\mathcal{E}_0=m_ec\omega_p/e$ is the cold nonrelativistic wavebreaking field, the maximum sustainable longitudinal electric field in a plasma in 1D, and we are considering linear polarization here. From this we can see that dispersion causes the wakefield to change in time and that GDD can be used to compensate for that. TOD and FOD can be explored similarly in a perturbative fashion, but in real life we are not restricted to perturbative spectral components and so must turn to simulations.

\section{Bayesian Optimization with Gaussian Processes}
\label{part2}

Optimization in laser-plasma acceleration is inherently challenging due to the high dimensionality of the parameter space, non-linear effects, and the high cost of simulations and experimental evaluations. Traditional grid searches or manual parameter scans quickly become infeasible in such regimes. To address this, Bayesian optimization (BO) offers a powerful framework to efficiently explore complex parameter spaces with limited evaluations.\cite{frazier2018}

BO is a surrogate-based global optimization method that probabilistically models an unknown objective function using a Gaussian process (GP). GP regression produces a posterior probability distribution where each $f(x)$ is normally distributed with mean $\mu(x)$ and variance $\sigma^2(x)$, updated after $n$ observations. The mean can be interpreted as a point estimate of $f(x)$, while the variance quantifies our uncertainty. At its core, a GP defines a distribution over functions characterized by a mean function $\mu(x)$ and a covariance kernel $k(x,x')$:
$$f(x) \sim GP[\mu(x),k(x,x')].$$
GPs are non-parametric, meaning they do not assume a fixed functional form but instead infer correlations directly from data. A key advantage of GPs over other machine learning methods such as neural networks or random forests is their built-in uncertainty quantification—the posterior variance explicitly indicates how uncertain the model is at any point in parameter space.

Given a set of observations, the GP posterior provides both a prediction and an uncertainty estimate at the new query points. At each iteration, the GP is used to construct an acquisition function $\alpha(x)$ that balances exploration (sampling in more uncertain regions) and exploitation (sampling near known high-performing points)  efficiently locate the global optimum. For this work, we employ the standard acquisition function of Expected Improvement (EI):
$${EI(x)}= [\mu(x)-f_\text{best}]\Phi(Z)+\sigma(x)\phi(Z),  \ \ \ \ \ \ Z=\frac{\mu(x)-f_\text{best}}{\sigma(x)},$$
again, where for a predictive distribution for $f(x)$, characterized by mean $\mu(x)$ and variance $\sigma^2(x)$, $f_\text{best}$ is the best observed value thus far and $\Phi$ and $\phi$ are the cumulative and probability distribution functions, respectively. The next evaluation point maximizes this acquisition function: $x_\text{next} = \arg\max_x \alpha(x)$. $EI$ trades off exploitation (large $\mu- f_\text{best}$) and exploration (large $\sigma$), naturally directing sampling towards regions that are promising or uncertain. While GP regression scales as $O(n^3)$ with the number of observations, this is acceptable for expensive black-box functions where $n$ typically remains below a few hundred evaluations

In this work, we apply BO to identify optimal spectral phase profiles and plasma configurations that maximize key LWFA metrics. Each configuration is evaluated through WarpX simulations that return scalar figures of merit. The GP surrogate interpolates smoothly in low-data regimes, which is critical given the computational cost of full LWFA simulations. Here we used the Scikit-Learn Bayesian optimization framework in addition to the Surrogate Modeling Toolbox.\cite{saves2024smt} For deeper mathematical treatments, we refer readers to comprehensive reviews on Gaussian processes and Bayesian optimization.\cite{frazier2018,Roussel2024Bayesian}

\section{Simulation Setup}
\label{setup}

In this study, we used the WarpX particle-in-cell (PIC) code to model laser wakefield acceleration. WarpX is a highly parallelized and optimized PIC code that can run on GPU and multi-core CPUs and includes load balancing capabilities.\cite{Fedeli2022} These simulations were run on the LLNL Lassen supercomputer and Perlmutter supercomputer at NERSC (LBNL). In addition, WarpX includes many physics and computational packages. Here we took advantage of the boosted-frame capability that allows us to relax the temporal resolution requirements of the simulation given that distances are scaled by the relativistic Lorentz factor $\gamma = 1/\sqrt{1 - v^2/c^2}$. The computational gain comes from the fact that in the boosted frame the laser is redshifted. The scale imbalance that results in long simulations in the lab frame is  reduced (the plasma is shorter and the laser wavelength is longer). In addition, we use the LASY laser package, which allows laser field and phase profiles to be inputted into the WarpX particle-in-cell code, as well as forward and backward propagate the laser pulse so that it can be properly injected into the WarpX simulation at the appropriate focus point.

In this study, our simulations have the following numerical setup. The simulation bounds are $[-256,0]\times[0,128] \mu m$ with a grid discretization of $[10,240] \times [128]$ for the longitudinal and horizontal coordinates, which means spatial resolutions of $0.025 \ \mu m$ and $1 \ \mu m$, respectively. The laser wavelength, a free variable in this study, is approximately 800 nm, meaning that we have 32 grid cells per laser wavelength longitudinally, which was found to be sufficient after extensive convergence testing and calibration. We use the boosted frame moving window package with a gamma boost factor of $\gamma = 10$; further enhancement leads to numerical artefacts forming that are considered to be nonphysical. Damped boundary conditions are used on the longitudinal boundaries of the simulation box, which applies a damping factor to the electric and magnetic fields in the outer half of the guard cells, using a sine-squared profile. This is necessary since we used the Pseudo Spectral Analytical Time Domain (PSATD) module to solve our system in RZ coordinates. We use 3 azimuthal modes and a momentum-conserving algorithm for field gathering. Particle count is varied across the length of the simulation, with higher counts in the initial, injection region of the plasma column, $(\text{ppc}_r,\text{ppc}_\theta,\text{ppc}_z)=(3,14,3)$, which is then relaxed after the laser passes through the dopant region, after which there is little injection still occurring, i.e., $(2,8,2)$. This was necessary to have convergent results for the electron bunch primarily via ionization injection but also to reduce the overall computational cost of the simulation. 

\begin{figure*}[h!t]
  \centering
  \begin{subfigure}[b]{0.45\textwidth}
    \includegraphics[height=2.5in]{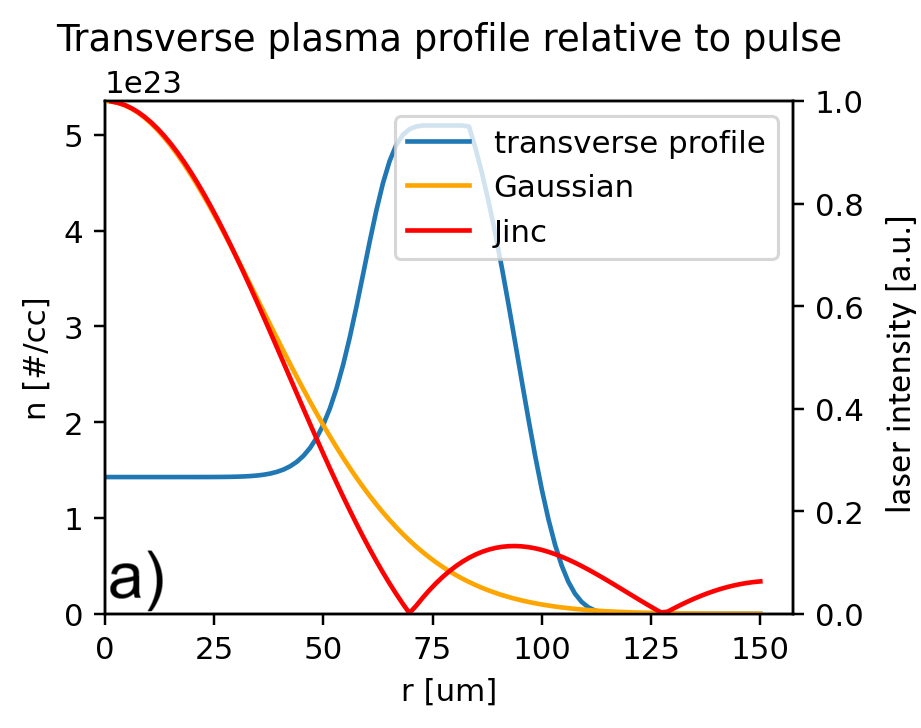}
  \end{subfigure}
  \hfill
  \begin{subfigure}[b]{0.45\textwidth}
    \includegraphics[height=2.5in]{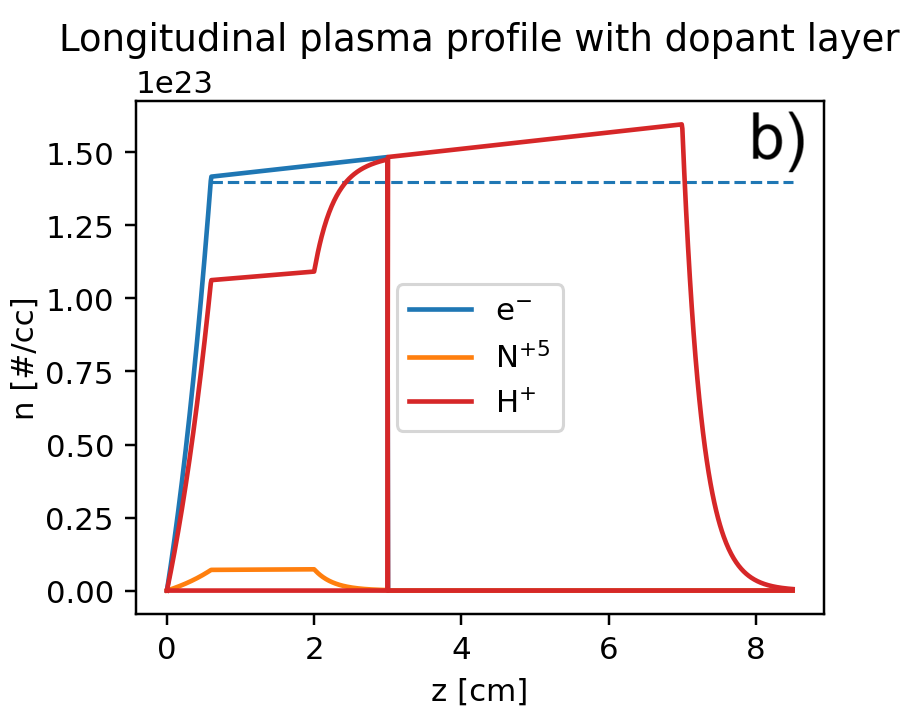}
  \end{subfigure}
  \caption{The a) transverse and b) longitudinal profiles used to describe the plasma column through which the laser drive propagates. The transverse profile is based on FLASH simulation results while the longitudinal profile includes a dopant layer, where the fully ionized hydrogen is doped with a few percent of nitrogen. In a) we compare the transverse profile to a Gaussian pulse with 50 $\mu m$ spot-size and a matched Jinc profile. Longitudinally, a density slope can be added to the plasma profile along its length, the propagation axis of the laser driver.}
  \label{fig:plasma}
\end{figure*}

The physical setup of our target problem is as follows. The laser driver is described by disjoint temporal and transverse functions, $a(r,\tau) = a_r(r)a_\tau(t,\text{GDD,TOD,FOD})$. Transversely it is described a Jinc function, i.e., $a_r(r)=\frac{2J_1(r/r_w)}{(r/r_w)}$, which describes an initially far-field top-hat laser profile as it reaches focus, as seen in Fig. \ref{fig:plasma}.a), where we compare it to a Gaussian transverse profile and a transverse slice of the plasma column. Here $r_w=r_0/2.66$, where $r_0=52 \ \mu m$ is the equivalent transverse pulse radius with respect to the Gaussian full-width at half maximum. Using LASY we can easily back-propagate a Jinc type laser so as to properly focus it into the target plasma. The temporal profile is described by 3 spectral components, i.e., GDD, TOD, and FOD, as previously described. The laser has a constant 30 J energy for all pulse shapes, although in experiments spectral shaping can effectively truncate parts of the spectrum so that the laser's intensity is less than what was theoretically predicted.


The plasma column through which the laser propagates is more complex than normally used. Here we used a custom plasma profile based on FLASH radiation hydrodynamics simulations that mimic the type of profile we expect to see given the HOFI plasma channel used in recent experiments.\cite{diaw2022,cook2020,cook2024,cook2025a} HOFI uses a preceding heater beam to generate the plasma profile to guide the drive laser over long distances. An example plasma profile can be seen in Fig. \ref{fig:plasma}.a). In addition, we have included longitudinal structure to the plasma channel to help optimize LWFA outputsdephasing. We typically have a plasma channel length of 60 cm in our simulations, although significant dephasing and depletion can begin as early as around 30 cm. The first few centimeters of the plasma is composed of a nitrogen-doped hydrogen gas, where the nitrogen is a few percent of the channel density. The hydrogen is fully ionized and the nitrogen is partially ionized to $N^{5+}$. As the laser propagates through the dopant layer it liberates $N^{6+}$ and $N^{7+}$ electrons, which constitute the primary contributions to the acceleration electron bunch, few background electrons feed into the final electron bunch. WarpX uses the Ammosov-Delone-Krainov (ADK) tunnel-ionization model. In addition, we allow for the plasma profile to have a density taper, where the density may increase linearly along the propagation axis $z$, which can be used to compensate for dephasing.\cite{rittershofer2010,djordjevic2019} A visualization of such a profile can be found in Fig. \ref{fig:plasma}.b).

\section{Enhancing LWFA yields with Bayesian Optimization}
\label{part5} 

In this study, we used Bayesian optimization with Gaussian processes to enhance scalar metrics of the accelerated electron bunch. First, this was done strictly with respect to the spectral components of the laser, i.e., GDD, TOD, and FOD, and then optimization was extended to laser wavelength, injection distance, and plasma profile characteristics. We used the Mat\'{e}rn covariance function as our kernel, which can better learn sharper gradients in parameter space than the standard Radial Basis Function most often used. The distance metric is unity for all inputs as we normalize within the minimum-maximum bounds of the sampling space.

To clarify, with respect to the laser driver, our free parameters are the central wavelength at which spectral modulation is applied, $\lambda_c$, GDD, TOD, FOD, and injection distance into the target $z_\text{inj}$. The pulse radius $r_0$ is not a free parameter in this study given experimental constraints on the modeled BELLA PW laser but could be in future studies. For the plasma profile, we have three tuning parameters, the length of the dopant region $\Delta z$, the dopant fraction of N in the background H plasma $f_\text{dop}$, and the slope of the longitudinal plasma profile $L_\text{slope}$. The ranges and bounds for the parameters sampled during optimization can be found in Table \ref{tab:bounds}.

\begin{table}
\caption{\label{tab:bounds} The bounds within which we performed BO of LWFA simulations. The baseline values correspond to the properties used for the Gaussian reference pulse as well as the first iteration of BO when only spectral parameters are varied. *For the initial BO run where we considered only spectral shaping, TOD and FOD were limited to bounds of [-12,000, 12,000] $fs^3$ and [-120,000, 120,000] $fs^4$. }
\begin{ruledtabular}
\begin{tabular}{|c|c|c|c|}
Variable & Min. & Max. & Base \\
\hline
GDD [$fs^2$] & -3,000 & 3,000 & - \\
TOD [$fs^3$]* & -24,000 & 24,000 & - \\
FOD [$fs^4$]* & -240,000 & 240,000 & - \\
$\lambda_c$ [nm] & 775 & 825 & 801 \\
$z_\text{inj}$ [cm] & 0.4 & 3.2 & 1 \\
$\Delta z$ [cm] & 0.4 & 6.2 & 2 \\
$f_\text{dop}$ [\%] & 2 & 22 & 5 \\
$L_\text{slope}$ [m$^{-1}$] & 0 & 20 & 0 \\
\end{tabular}
\end{ruledtabular}
\end{table}

\subsection{Reference result: Gaussian pulse}

The standard way to model LWFA in the modeling community is to use a Gaussian laser pulse, both transversely and longitudinally. We have already enhanced the fidelity of our result by modeling the transverse profile with a Jinc profile, but typically in most studies that do that a Gaussian longitudinal profile is still used despite not matching experiments. We will provide here a reference result to which we will compare our BO-enhanced results. In this case, we based our Gaussian mode on what is typically used to model the BELLA PW, that is 52 $\mu m$, $\tau=92 \ fs$, and an energy of 30 J, which gives an approximate intensity of $I_0=4.53\times10^{16}$ W/cm$^2$ or normalized intensity of $a_0=1.46$, placing this design in an intermediate place between the quasi-linear and blowout regimes. Using a Gaussian pulse, we achieve $E_m = 9.36$ GeV, $Q = 10.31$ pC, and $\sigma_E = $ 2.96 \% after 47.85 cm. WarpX simulations using these parameters were calibrated to and closely matched preliminary INF\&RNO simulations.\cite{benedetti2018}


\subsection{Grid scan with respect to spectral shaping}

\begin{figure*}[h!t]
    \begin{subfigure}{\textwidth}
        \centering
        \includegraphics[width=0.99\textwidth]{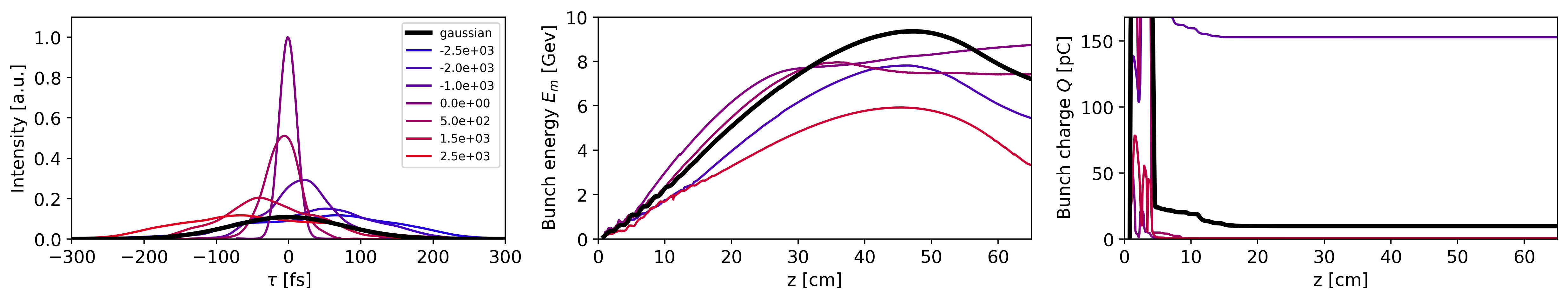}
        \caption{Grid scan of GDD from -2,500 to 2,500 $fs^{-2}$.}
        \label{fig:sub1}
    \end{subfigure}
    \begin{subfigure}{\linewidth}
        \centering
        \includegraphics[width=0.99\textwidth]{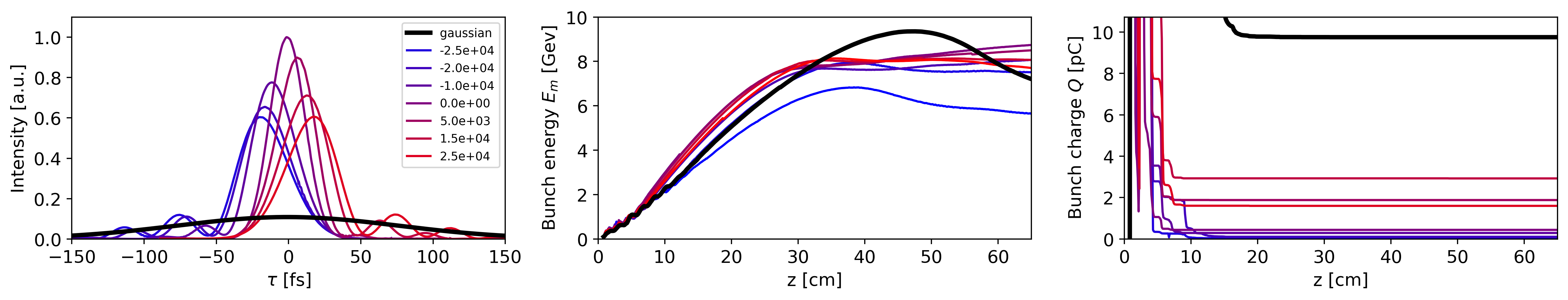}
        \caption{Grid scan of TOD from -25,000 to 25,000 $fs^{-3}$.}
        \label{fig:sub2}
    \end{subfigure}
    \begin{subfigure}{\linewidth}
        \centering
        \includegraphics[width=0.99\textwidth]{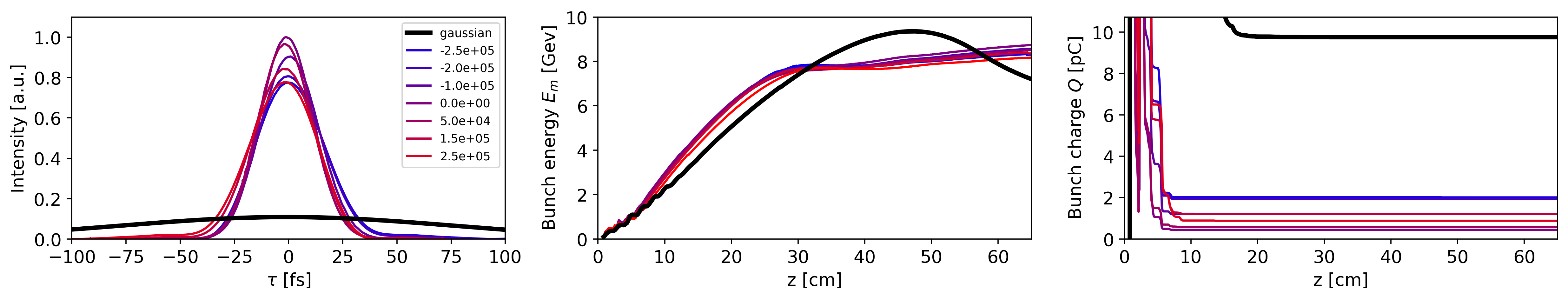}
        \caption{Grid scan of FOD from -250,000 to 250,000 $fs^{-4}$.}
        \label{fig:sub3}
    \end{subfigure}
\caption{\label{fig:gridscan} Grid scans of GDD, TOD, and FOD while keeping other parameters constant relative to the baseline listed in Table. \ref{tab:bounds}. Plotted are the laser intensities normalized to the transform-limited pulse (GDD=TOD=FOD=0), the mean energy of the accelerated bunch $E_m$, and the bunch charge $Q$. The reference Gaussian pulse and its results are shown in black.}
\end{figure*}

To elucidate how spectral shaping affects the drive laser and baseline metrics of electron bunch mean energy and charge, we present grid scans of GDD, TOD and FOD in Fig. \ref{fig:gridscan}. In the first column, we plot the laser pulse shapes relative to spectral shaping and normalized them to the transform-limited pulse shape and compared them to an idealized pulse with a Gaussian longitudinal profile with a transverse Jinc profile. In the second column, we show the mean energy of the accelerated electron bunch and in the third column the amount of charge in the bunch. We can see that the Gaussian performs well overall, reaching the highest energy and then an intermediate level of charge. As we go higher in spectral content, even while increasing the bounds by an order of magnitude, we have a smaller effect on the laser pulse and accelerated bunch properties. However, the complexity is further extended so that the spectral components can mix, in particular GDD and TOD, thus further expanding the parameter space. Modifying multiple spectral components at once will allow us to exceed the performance of an idealized Gaussian pulse. To help us explore this large parameter space, we will use BO.

\subsection{Bayesian Optimization with respect to spectral shaping}
\label{subsecbo1}
The initial thrust of this study was to use spectral shaping, i.e., manipulating GDD, TOD, and FOD, to maximize the mean energy of the accelerated electron bunch, $E_m$. In this case and all subsequent cases, we only consider electrons with 0.1 GeV in energy or more. This study entailed several iterations of progressive complexity and breadth to find a result that achieved the target goals. Initially, we started by using BO to minimize a simple objective function of $f_\text{obj} = -E_m$. This proved to be insufficient because while $E_m$ was readily enhanced to $\gtrsim$ 20 GeV, it was done so at severe loss of charge, i.e., $\ll$ 1 pC levels.

\begin{figure}[h]
    \begin{subfigure}{\columnwidth}
        \centering
        \includegraphics[width=0.99\columnwidth]{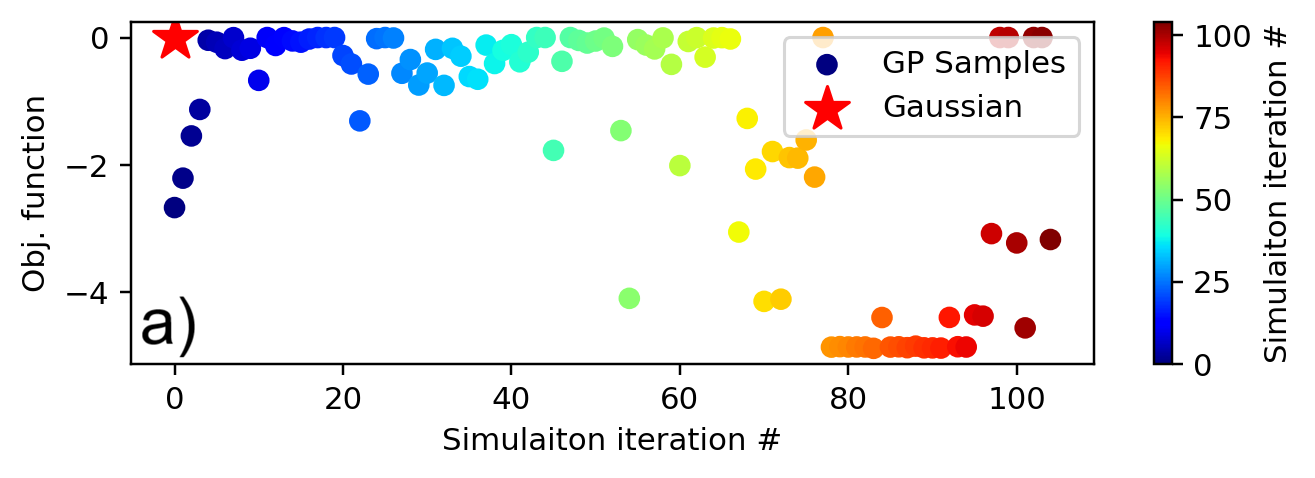}
    \end{subfigure}
    \begin{subfigure}{\linewidth}
        \centering
        \includegraphics[width=0.99\columnwidth]{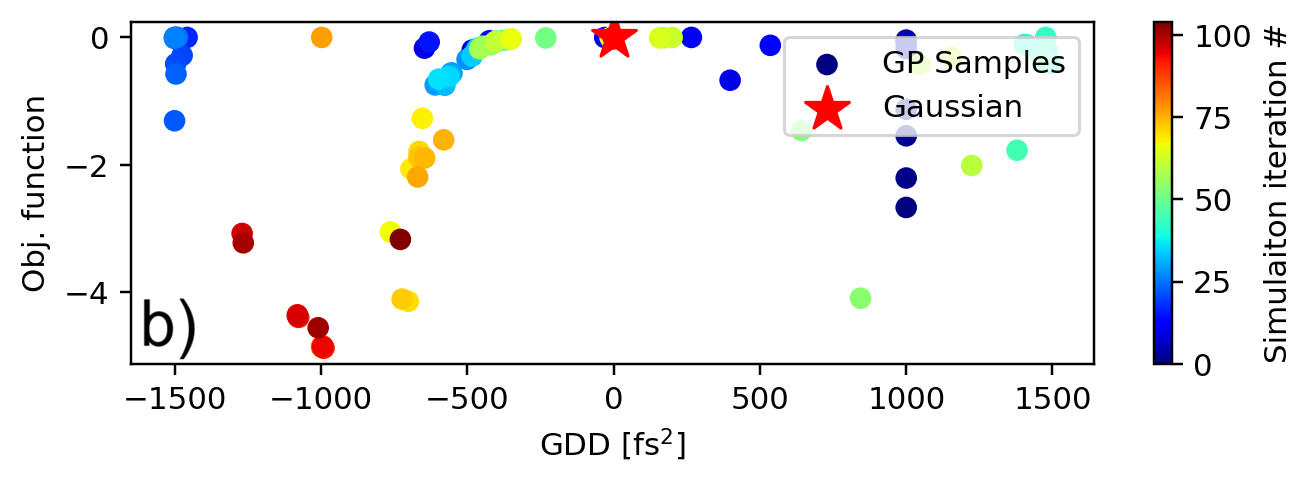}
    \end{subfigure}
    \begin{subfigure}{\linewidth}
        \centering
        \includegraphics[width=0.99\columnwidth]{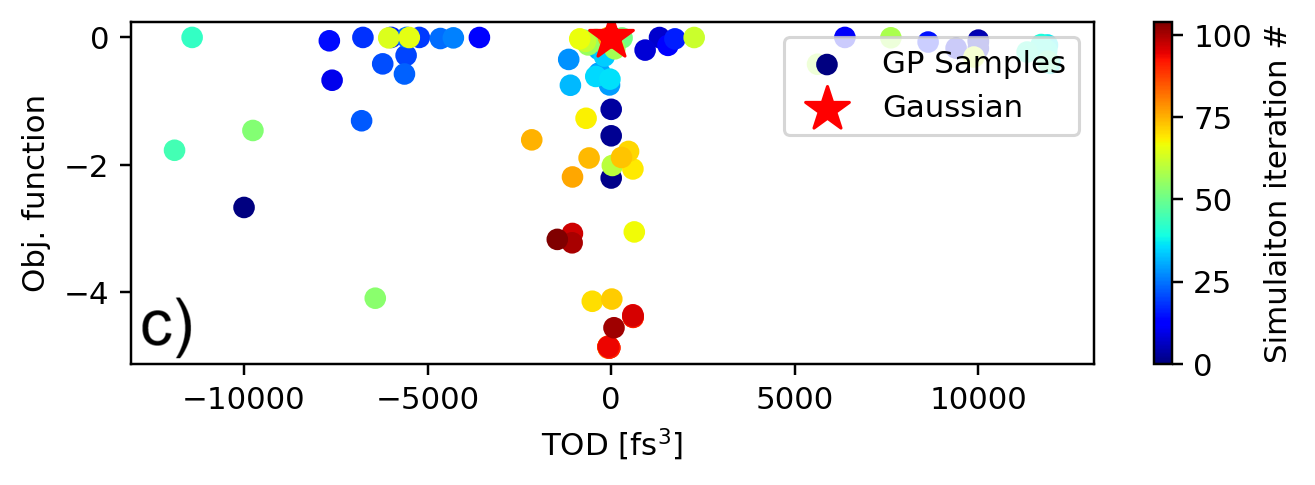}
    \end{subfigure}
    \begin{subfigure}{\linewidth}
        \centering
        \includegraphics[width=0.99\columnwidth]{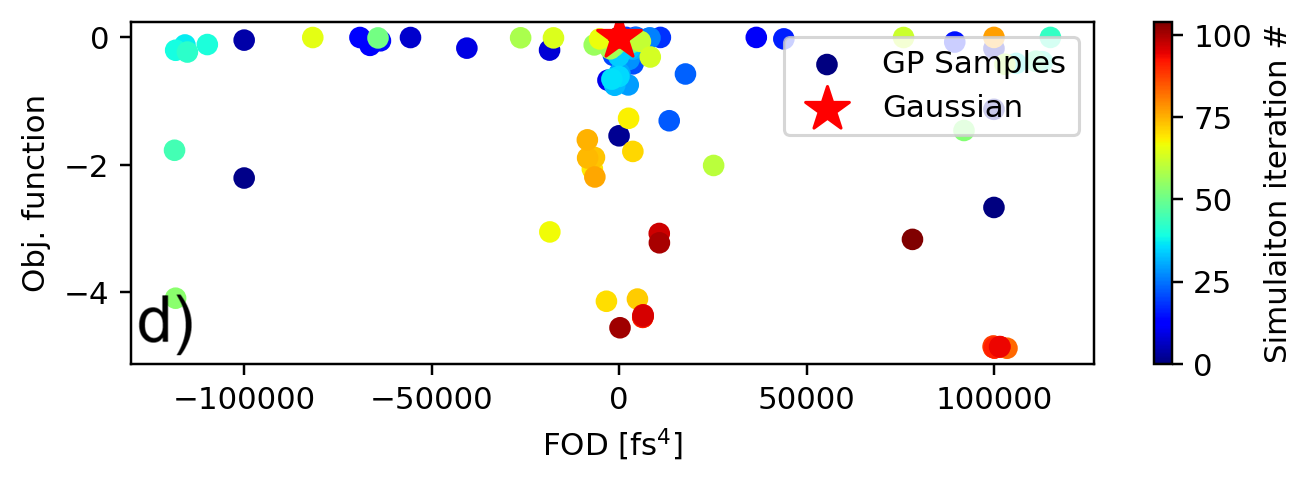}
    \end{subfigure}
\caption{\label{fig:inital_bo} Initial Bayesian optimization where we show the simulation iteration number in a) with respect to the spectral components b) GDD, c) TOD, and d) FOD.}
\end{figure}

To compensate for this, we implemented a composite objective function on which BO operated of the following form:
\begin{equation}
    f_\text{obj} = a-\log_{10}(C_Q|Q|+b_Q)\times(C_E E_m+b_E)/(C_\sigma\sigma_E+b_\sigma),
    \label{objfunc}
\end{equation}
where again $Q$ is the charge in the electron bunch, $\sigma_E=\Delta E/E_m$ is the energy spread in the accelerated electrons and $\Delta E$ is defined as the standard deviation about the electron energy peak, and $a$, $b$, and $C$ are hyperparameters used to adjust the predictive performance of the GP so that it best reproduces the data prior to selecting new points to sample in parameter space. For this initial investigation, the hyperparameters were set to $a=0$, $C_Q=10^{14}$, and $C_E=C_\sigma=b_Q=b_E=b_\sigma=1$. These were manually tuned during the study. Similarly, this initial study we only simulated the interaction up to 40 cm, while dephasing can occur as late as 60 cm in the following studies. The initial BO scan can be found in Fig.\ref{fig:inital_bo}, where we have plotted a) the simulation iteration \# as the GP explores the parameter space, minimizing uncertainty and the objective function, and then b) GDD, c) TOD, and d) FOD parameters during optimization.

 After approximately 80 iterations, BO found a new optimum, \#83, which it subsequently reiterated to minimize uncertainty. The path of optimization is particularly clear with respect to GDD, as BO marches through the parameter space until it settles on approximately (-991 fs$^{-2}$,-79 fs$^{-3}$, 103,473 fs$^{-4}$). In this case, TOD is a relatively minor contribution, with, although there is a noticeably significant performer, iteration \#54, with (-844 fs$^{-2}$,-6,425 fs$^{-3}$, -118,290 fs$^{-4}$), which has a more significant contribution to TOD. 

\begin{figure*}[t]
    \centering
    \begin{subfigure}{0.23\textwidth}
        \centering
        \includegraphics[height=0.85in]{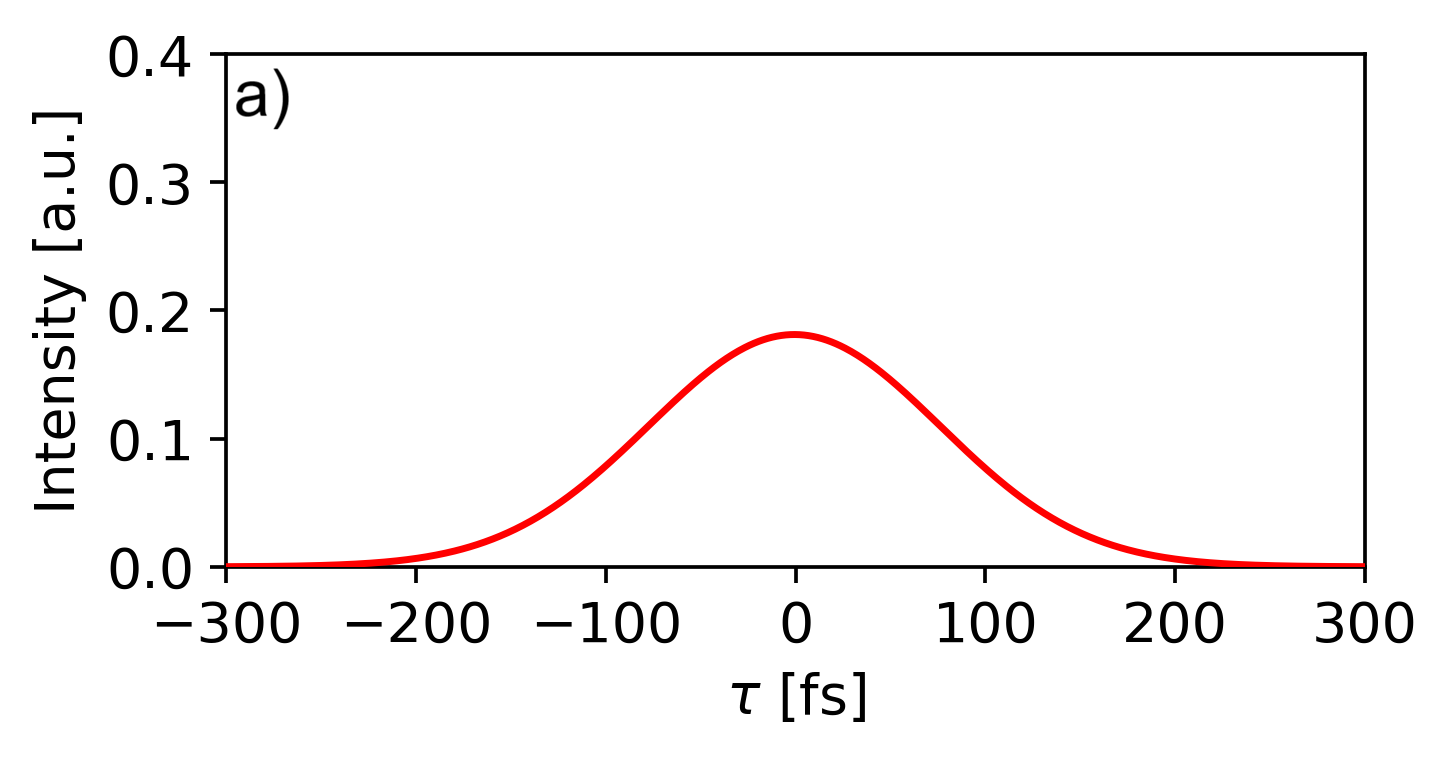}
    \end{subfigure}%
    \begin{subfigure}{0.23\textwidth}
        \centering
        \includegraphics[height=0.85in]{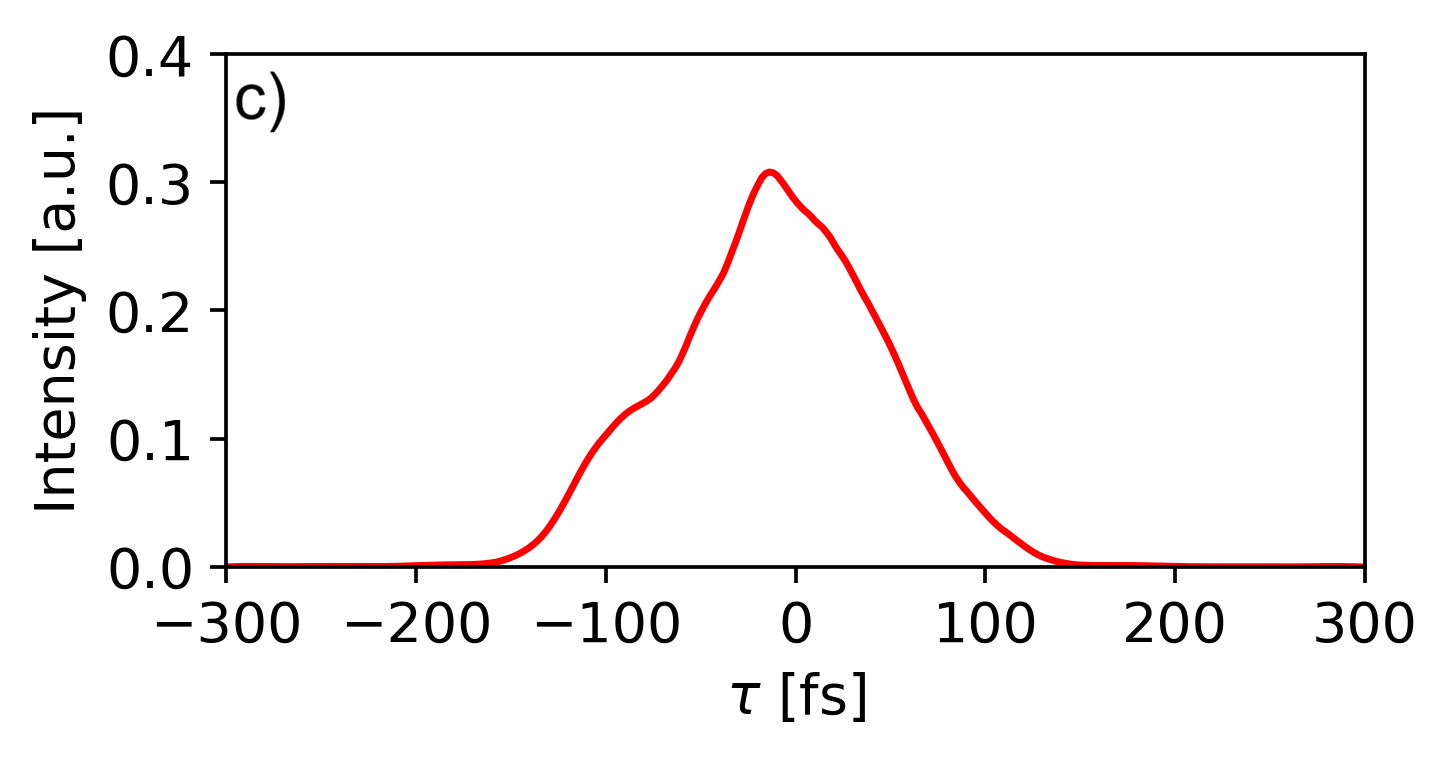}
    \end{subfigure}%
    \begin{subfigure}{0.23\textwidth}
        \centering
        \includegraphics[height=0.85in]{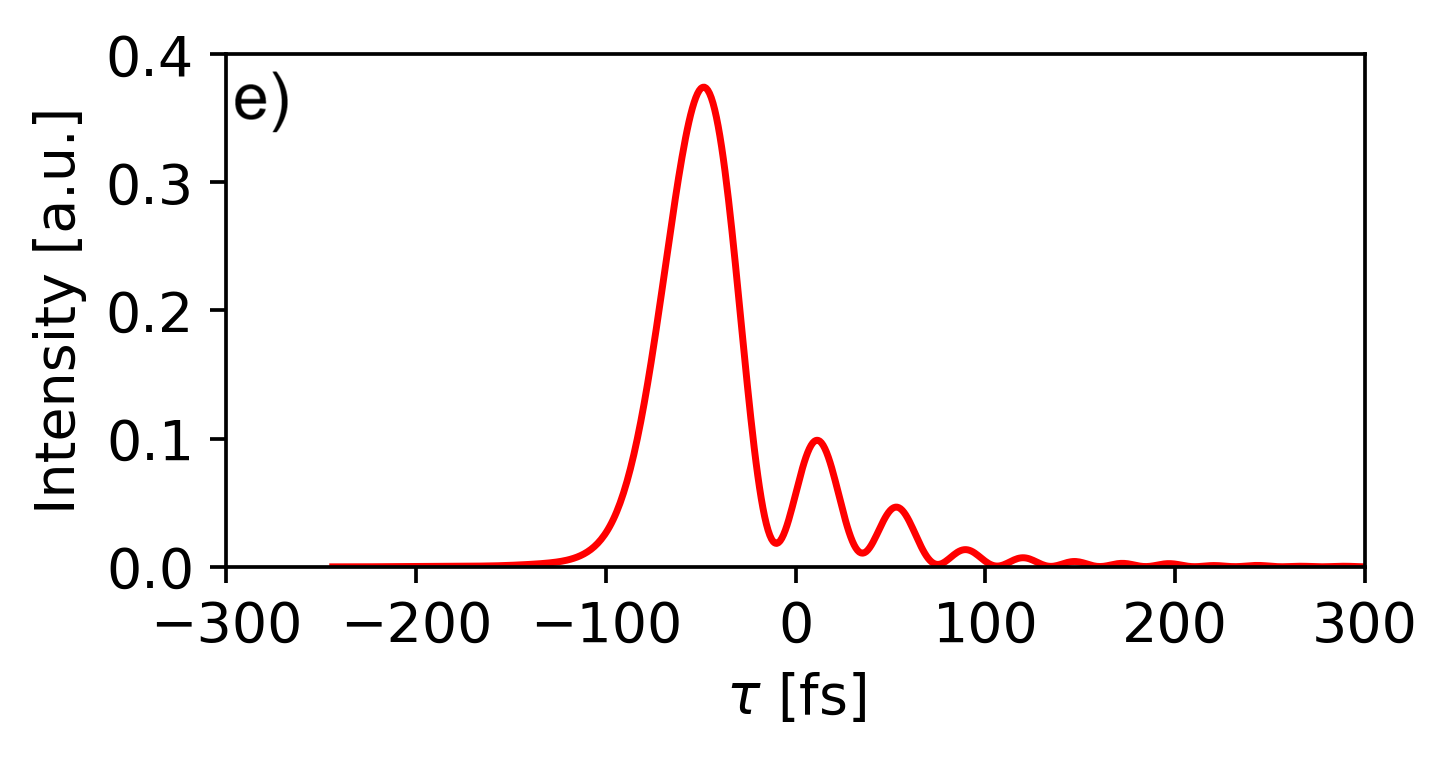}
    \end{subfigure}%
    \begin{subfigure}{0.23\textwidth}
        \centering
        \includegraphics[height=0.85in]{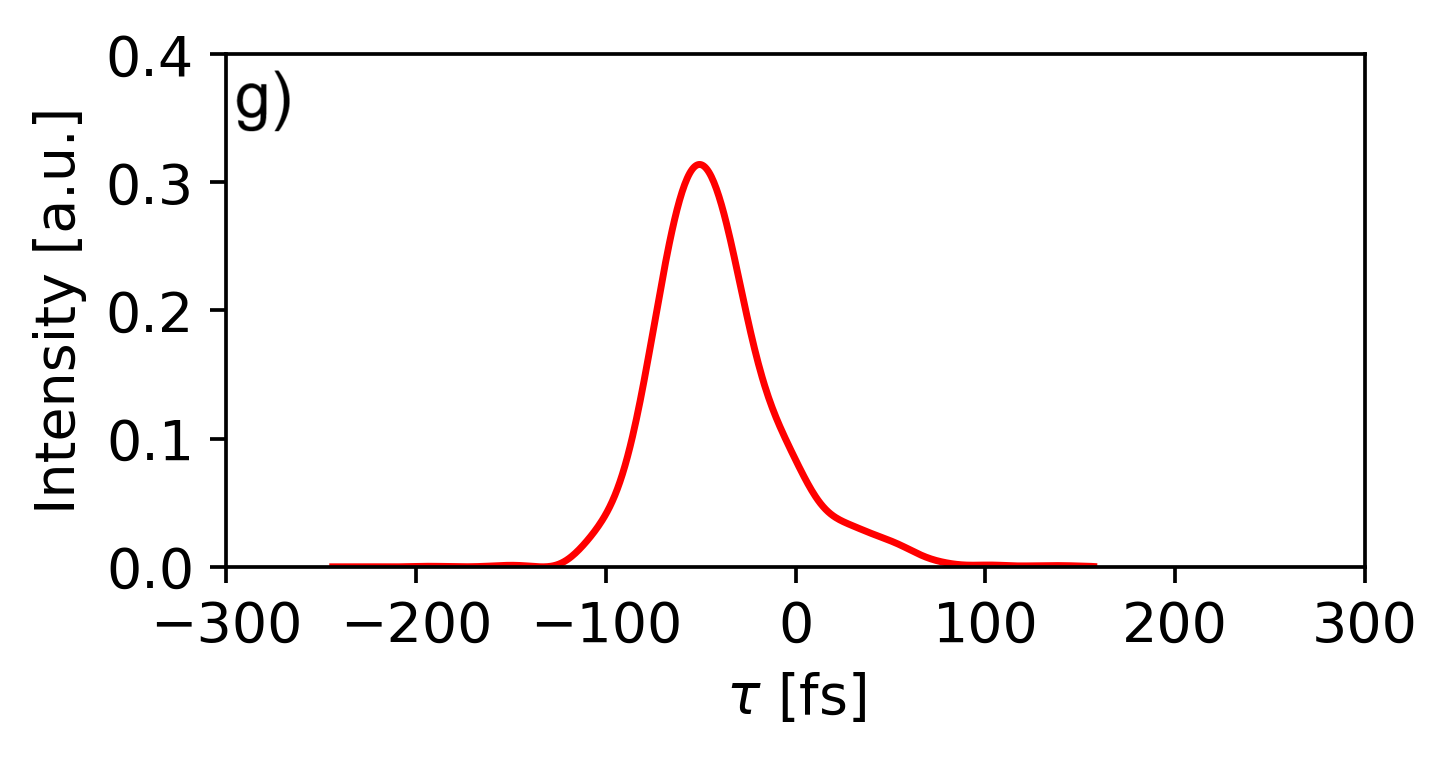}
    \end{subfigure}%
    \\[1ex]
    \begin{subfigure}{0.23\textwidth}
        \centering
        \includegraphics[height=1.37in]{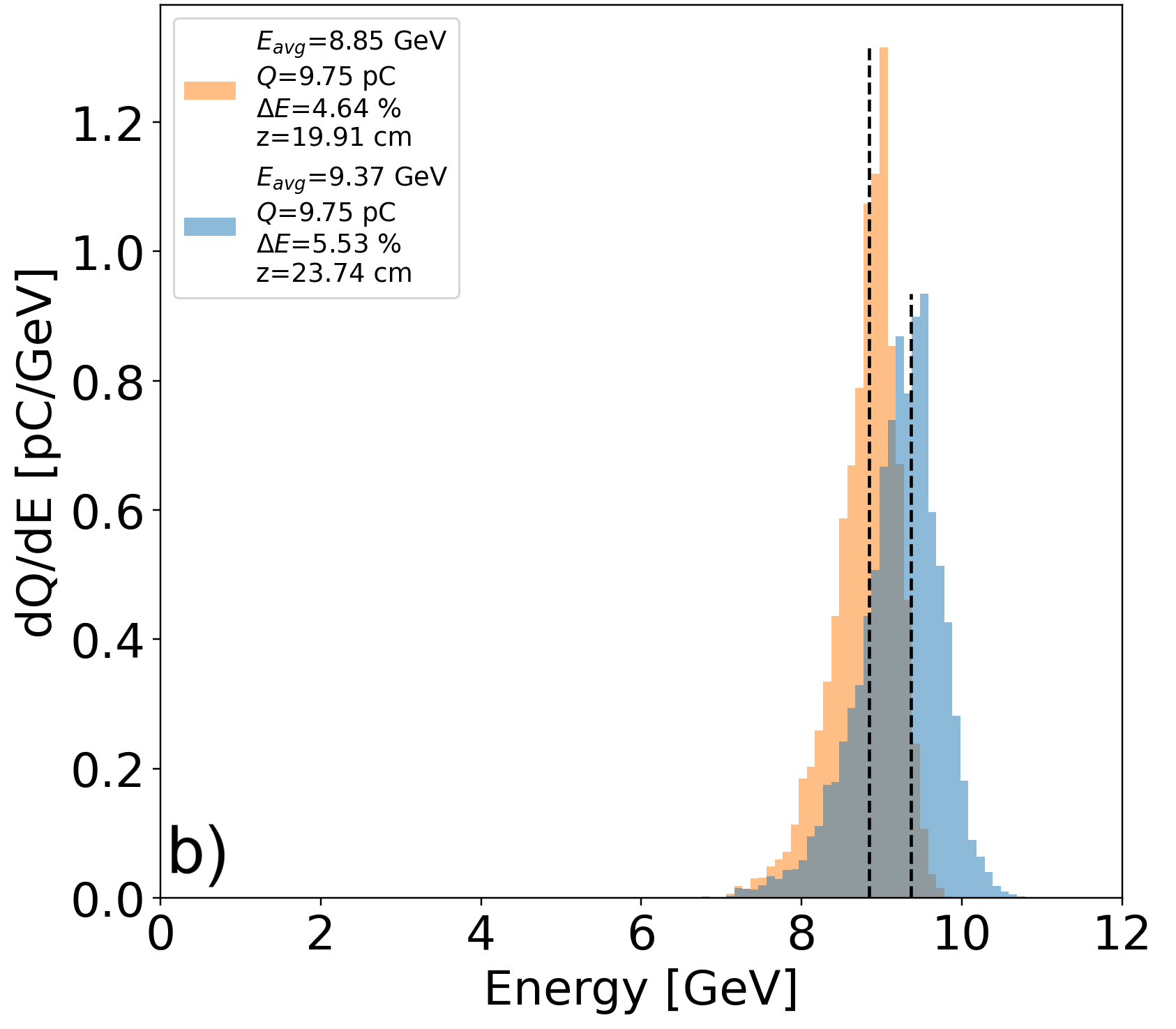}
    \end{subfigure}%
    \begin{subfigure}{0.23\textwidth}
        \centering
        \includegraphics[height=1.37in]{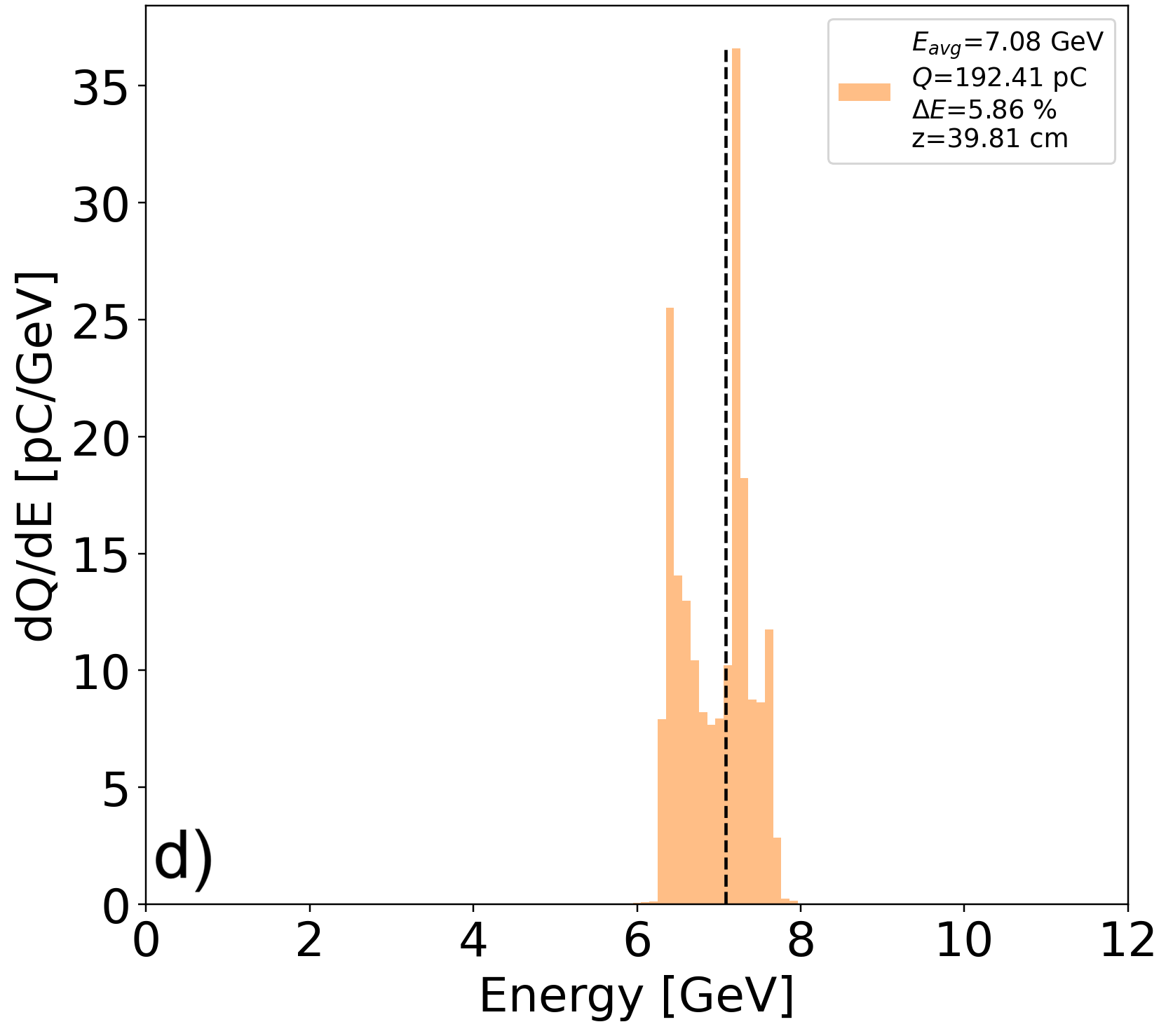}
    \end{subfigure}%
    \begin{subfigure}{0.23\textwidth}
        \centering
        \includegraphics[height=1.37in]{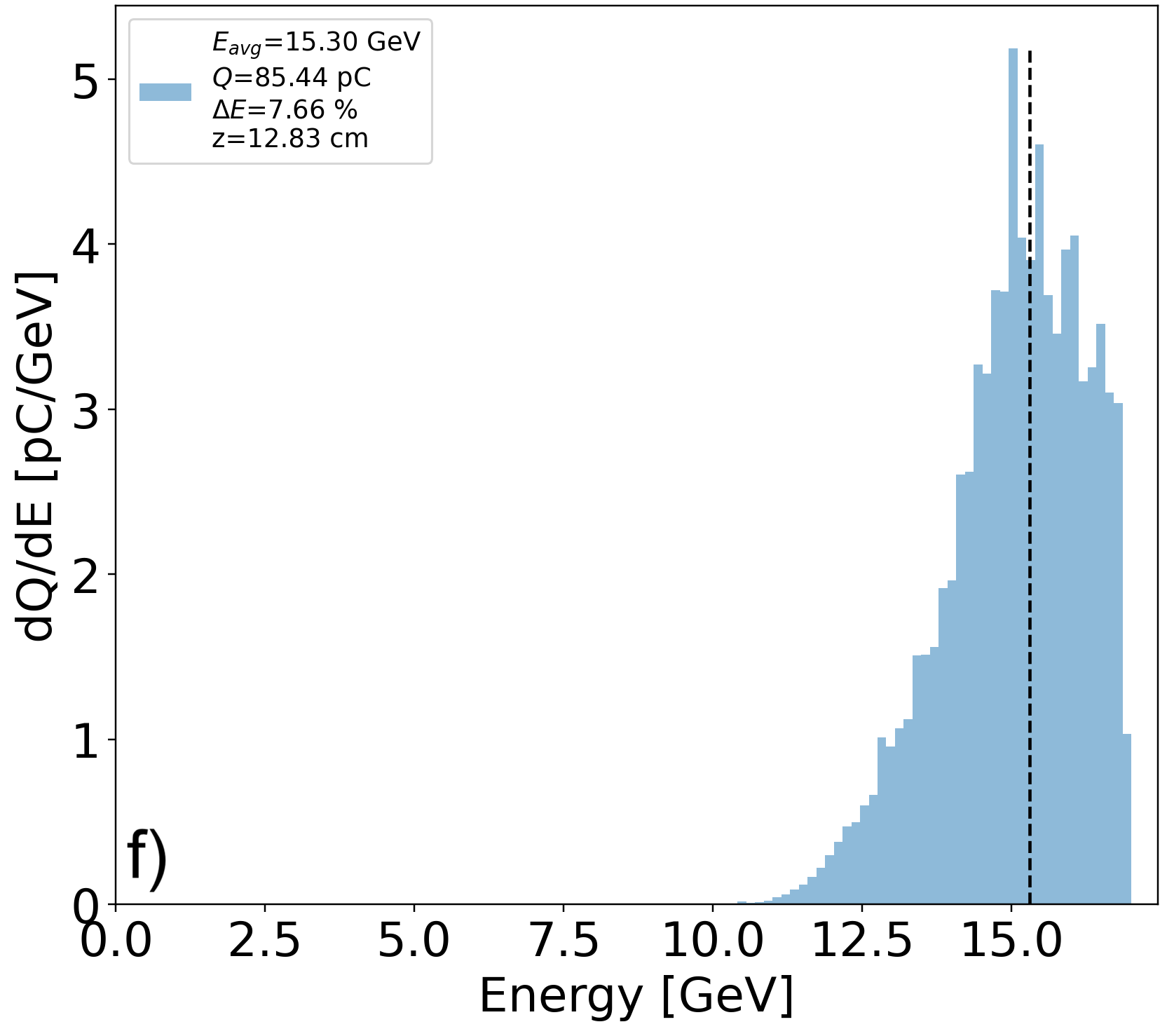}
    \end{subfigure}%
    \begin{subfigure}{0.23\textwidth}
        \centering
        \includegraphics[height=1.37in]{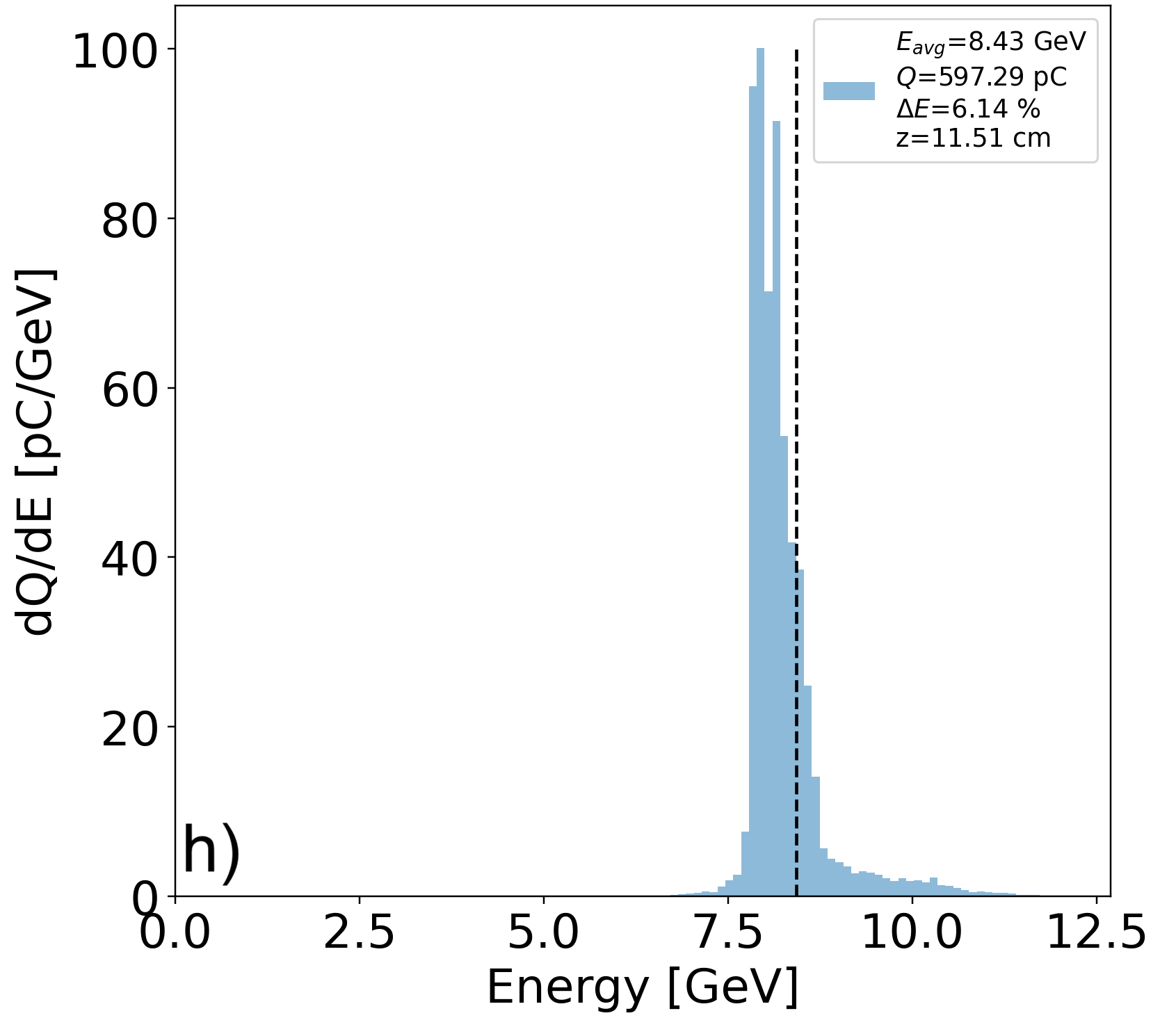}
    \end{subfigure}%
    

    \caption{\label{fig:laser_hist} The laser profiles and associated electron energy spectra for a,b) the Gaussian with respect to the simulation run distance for BO scan \#1 (orange, 40 cm) and \#2 (blue, 75 cm), c,d) the peak performer for BO scan \#1, e,f) the high energy (HE) and g,h) the high charge (HQ) cases from BO scan \#2. The laser profiles are normalized to the transform-limited intensity.
    }
\end{figure*}

Concentrating on the current highest performer with respect to the objective function and just the spectral components, we present the pulse shape and associated energy spectrum of the electron bunch in Fig. \ref{fig:laser_hist}.b). The laser pulse in Fig. \ref{fig:laser_hist}.b) has a degree of kurtosis, as well as a prepulse bump prior to the peak. The energy spectrum in Fig. \ref{fig:laser_hist}.b) has a lower energy than the Gaussian case before, 7.1 GeV versus 9.36 GeV, but much more charge, 193.7 pC versus 10.3 pC. Likewise, despite the unusual double-peaked energy spectrum, the energy spread is not significantly worse than that of the Gaussian driver, 5.81\% versus 2.96\%.



\subsection{Bayesian Optimization with respect to spectral shaping, laser, and plasma parameters}

Although we were able to enhance the charge content of the accelerated bunch with BO, we did so at the expense of mean energy. To see if we can further optimize the system, we extend our GP parameter space from 3 to 8 dimensions as well as extend the range of TOD and FOD per Table \ref{tab:bounds}, where we now include the central wavelength $\lambda_c [nm]$, injection distance $z_\text{inj} [cm]$, dopant layer thickness $\Delta z [cm]$, dopant fraction $f_\text{dop} [\%]$, and longitudinal density slope $L_\text{slope} [m^{-1}]$. 

\begin{figure}[h!t]
        \centering
\includegraphics[width=0.99\columnwidth]{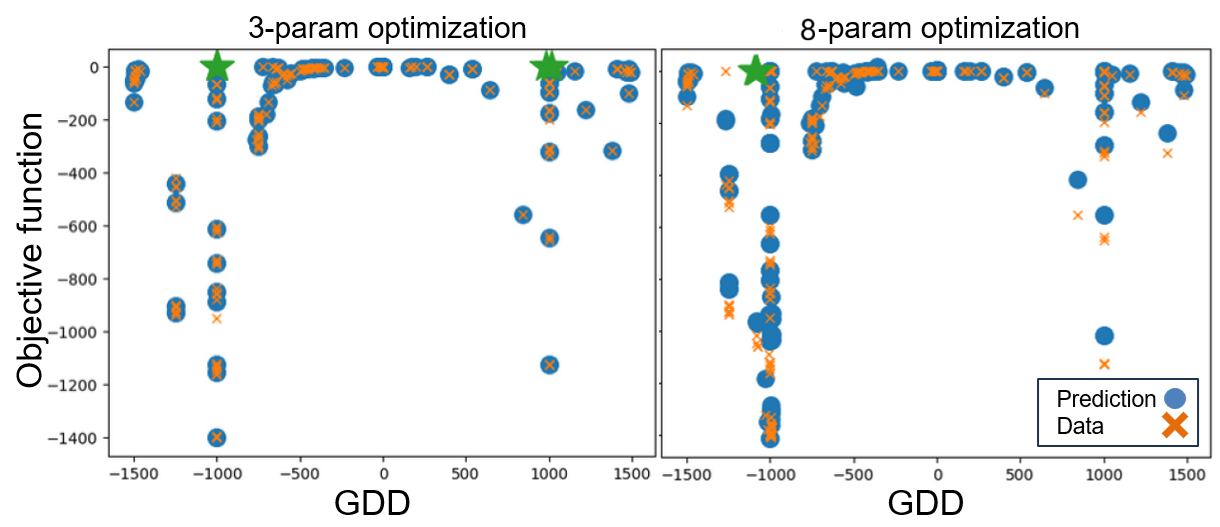}
\caption{\label{fig:comp_3_7} Comparison of 3 parameter training on initial dataset versus 8 parameter training on a new dataset, including the sample points (circles) and the GP surrogate prediction (crosses). }
\end{figure}

One of the disadvantages of GPs for regression or optimization is that they do not scale well with larger dimensions. For example, in Fig. \ref{fig:comp_3_7}, when we increase the parameter space from 3 to 8 but use the same hyperparameters per Sec. \ref{subsecbo1}, the GP surrogate model is less capable of reproducing the training data without further calibration, thus the inclusions of hyperparameters in Eq. \ref{objfunc}. These were manually calibrated several times while generating and training new batches of simulations, where in the end we ran 193 simulations with coefficients $a=2$, $C_Q=10^{15}$, $C_E=C_\sigma=1$, and $b_Q=b_E=b_\sigma=5$. This was done in steps, initially starting with a gradient scan with respect to each parameter, i.e., 8 simulations, then subsequently individual simulations for the next 8 simulations, varying all parameters, batches of three simulations for the next 42, and then 130 simulations in batches of 10. A unique advantage of GPs is their ability to select and train on new points in parallel, unlike gradient-based methods. 

\begin{figure}[h!t]
        \centering
\includegraphics[width=0.99\columnwidth]{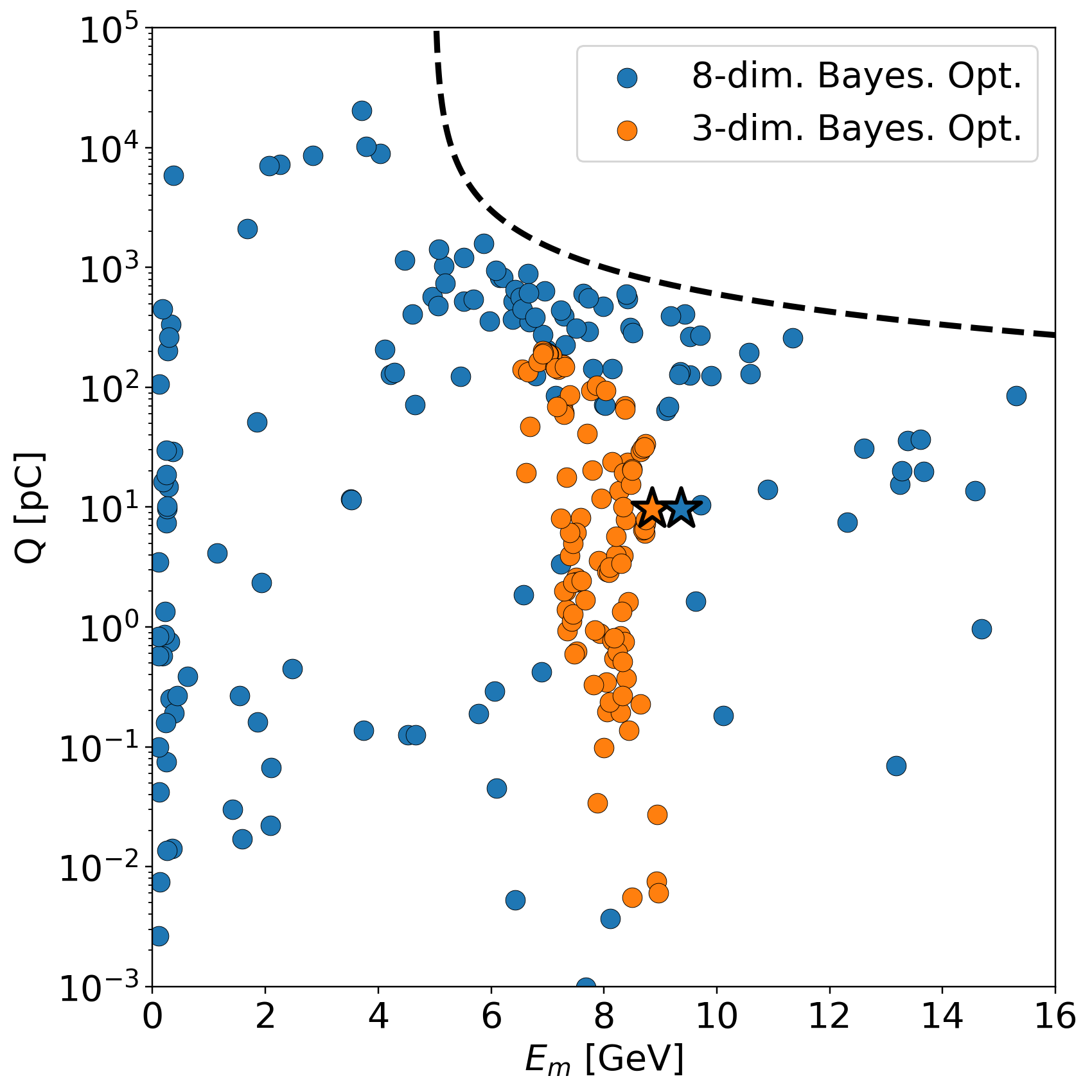}
\caption{\label{fig:pspace2} Parameter space sampling with respect to output metrics $E_m$ and $Q$. Presented are the initial sampling with respect to just spectral parameters GDD, TOD, and FOD (orange), and then the larger sampling with respect to also $\lambda_c$, $z_\text{inj}$, $\Delta z$, $f_\text{mix}$, $L_\text{slope}$ (blue). Plotted also is an approximate Pareto front (dashed). Reference Gaussian is presented as well, for the spectral-only, 3-dimensional BO case (orange star), and for the 8-dimensional BO case (blue star).}
\end{figure}

The parameter space sampled with respect to the output metrics is presented in Fig. \ref{fig:pspace2}, where we show the $E_m$ vs. $Q$ slice. We can see that the parameter space sampled with just spectral components (orange) is much smaller than that sampled using an 8-dimensional sampling space (blue), which is to be expected. The baseline Gaussian pulse is also plotted for both datasets (stars), where the simulation range was increased from 40 cm to 75 cm. Depicted is an estimate of the Pareto front (dashed), where we plot $Q (E_m-5)=3\times10^3$, which takes advantage of the simple intuition that the product $QE_m\approx const.$ is often approximately conserved in ideal circumstances \cite{tzoufras2008}.


\begin{figure}
    \centering

    
        \centering
        \includegraphics[width=\linewidth]{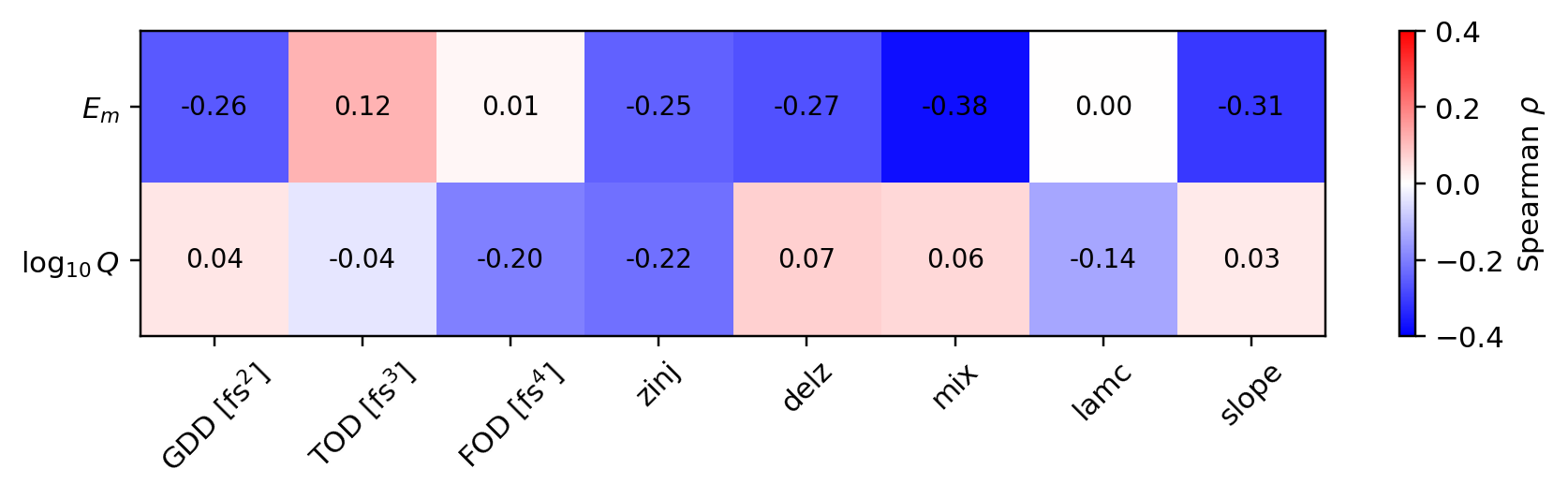}
    \caption{\label{fig:pspace_dep} $Q$ and $E_m$ variance with respect to the input parameters. A correlation matrix of inputs versus outputs with respect to Spearman rank correlation.
    }
\end{figure}

A matrix plot of the Spearman rank correlation is presented in Fig. \ref{fig:pspace_dep}. The Spearman rank correlation, $\rho=r\left(\text{rank}(X),\text{rank}(Y)\right)$ with $r = \text{cov}(X,Y)/\sigma_X\sigma_Y$, measures monotonic associations both linear and nonlinear, making it less sensitive to outliers and invariant to monotonic transformations, e.g., $\log_{10}(Q)$. The strongest correlation is between $f_\text{dop}$ and $E_m$, where the energy drops as we increase the mix fraction. This correlates to beamloading, wherein a higher electron bunch charge reduces the amplitude of the accelerating field, and thereby achieves lower energies. Based on this figure and additional analysis, it appears that FOD, $\lambda_c$, and injection distance $z_\text{inj}$ have little effect on $Q$ or $E_m$. Higher levels of dopant fraction, GDD, and $L_\text{slope}$ result in case examples with the highest values of $Q$ but lowest values of $E_m$. 


\section{High-performing, BO-enabled case examples}
\label{partexamples}
 
With only spectral shaping we could significantly increase the charge relative to the Gaussian case but not the mean energy significantly without losing charge in turn. By incorporating the laser wavelength, injection range, and channel plasma conditions into the BO loop, we found new optima for both $E_m$ and $Q$ without severely sacrificing one or the other. The optimization path was somewhat skewed by the objective function, thus the clustering of points near $Q\sim 200$ pC and $E_m\sim$ 7 GeV. We will investigate several high performers and how they differ from the baseline Gaussian case.

Based on the eight-dimensional objective function and its associated coefficients, the peak performer selected, which we designate the High Energy case (HE), has the following setup conditions found in Table \ref{tab:best1}. These can be compared to the baseline parameters that informed the 3-dimensional optimization scan and the reference Gaussian pulse case. Several of the HE parameters approach the boundaries of the initial sampling range, notably $\lambda_c$, FOD, and $z_\text{inj}$, which may suggest pushing those limits further. The HE metrics are $E_m= 15.3$ GeV, $Q = 85.4$ pC, and $\sigma_E = 7.7$ \%. The associated laser pulse and electron energy spectrum at peak $E_m$ can be found in Fig. \ref{fig:laser_hist}(e,f). Another notable case is the fourth highest performing simulation, the High Charge case (HQ), whose input parameters can be found in Table. \ref{tab:109} and the associated laser and energy spectrum in Fig. \ref{fig:laser_hist}(g,h). The output metrics for this simulation are $E_m= 8.43$ GeV, $Q = 597.3$ pC, and $\sigma_E = 6.1$ \%.

\begin{table}
\caption{\label{tab:best1} High energy (HE) case: the input parameters for the highest performing simulation per BO in the 8-dimensional optimization scan.}
\begin{tabular}{|c|c|c|c|}
\hline
GDD [$fs^2$] & TOD [$fs^3$] & FOD [$fs^4$] & $\lambda_c$ [nm] \\
-1.89e+03 & 1.95e+04 &2.34e+05 & 824 \\ 
\hline
\hline
$z_\text{inj}$ [cm] & $\Delta z$ [cm] & $f_\text{dop}$ [\%] & $L_\text{slope}$ [m$^{-1}$] \\
0.52 & 1.201 & 14.54 & 9.01 \\
\hline
\end{tabular}
\end{table}

\begin{table}
\caption{\label{tab:109} High charge (HQ) case: The input parameters for the fourth highest performing simulation per BO in the 8-dimensional optimization scan with particularly high charge.}
\begin{tabular}{|c|c|c|c|}
\hline
GDD [$fs^2$] & TOD [$fs^3$] & FOD [$fs^4$] & $\lambda_c$ [nm] \\
-1.26e+03 & 1.56e+04 &-1.47e+05 & 825 \\ 
\hline
\hline
$z_\text{inj}$ [cm] & $\Delta z$ [cm] & $f_\text{dop}$ [\%] & $L_\text{slope}$ [m$^{-1}$] \\
0.52 & 3.95 & 10.9 & 6.42 \\
\hline
\end{tabular}
\end{table}



\begin{figure}[h!t]
    \begin{subfigure}{\columnwidth}
        \centering
        \includegraphics[width=0.99\columnwidth]{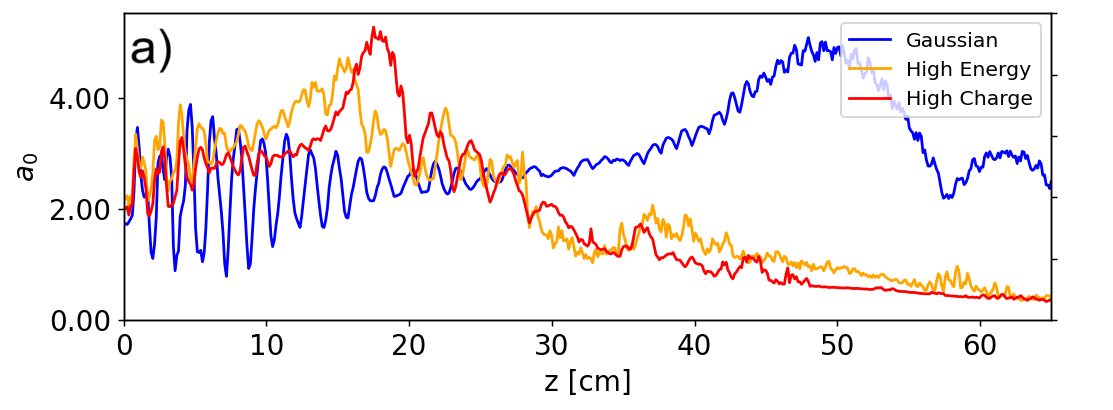}
    \end{subfigure}
    \begin{subfigure}{\columnwidth}
        \centering
        \includegraphics[width=0.99\columnwidth]{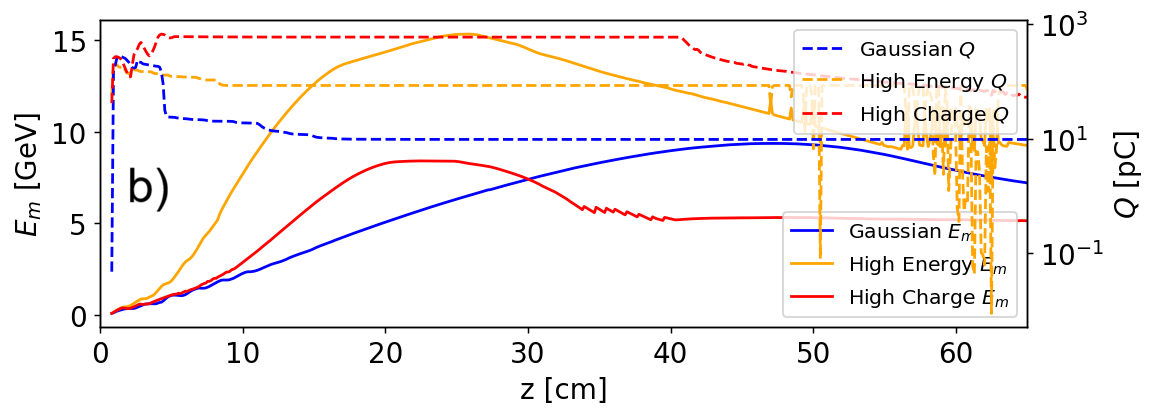}
    \end{subfigure}
    \begin{subfigure}{\columnwidth}
        \centering
        \includegraphics[width=0.99\columnwidth]{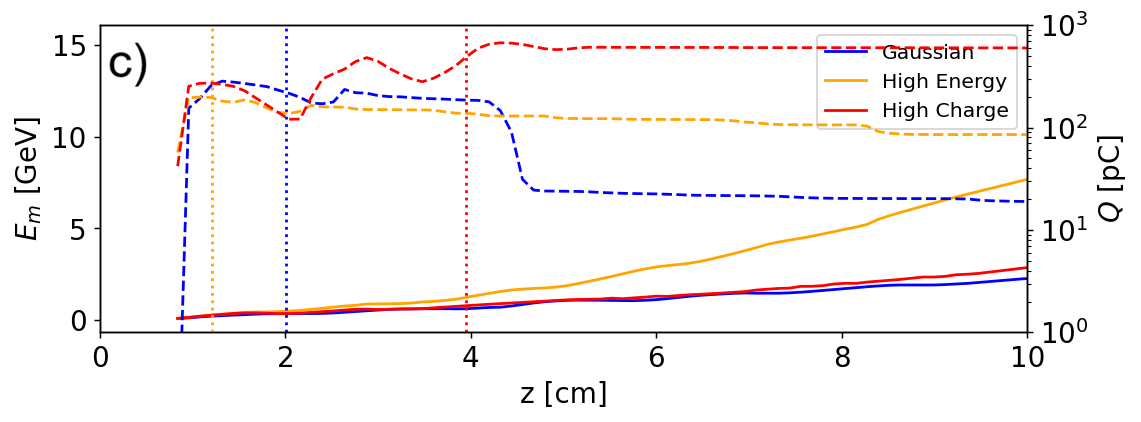}
    \end{subfigure}
\caption{\label{fig:a0_osc} a) Comparison of peak $a_0$ as a function of propagation distance for the baseline Gaussian (red), HE case (orange), and HQ case (red) and then b) $E_m$ and $Q$ as a function of time. In sub-figure c) we focus in on $E_m$ and $Q$ earlier in time, where the dopant region cutoff is illustrated (vertical lines). }
\end{figure}
\begin{figure}
    \begin{subfigure}{\columnwidth}
        \centering
        \includegraphics[width=0.99\columnwidth]{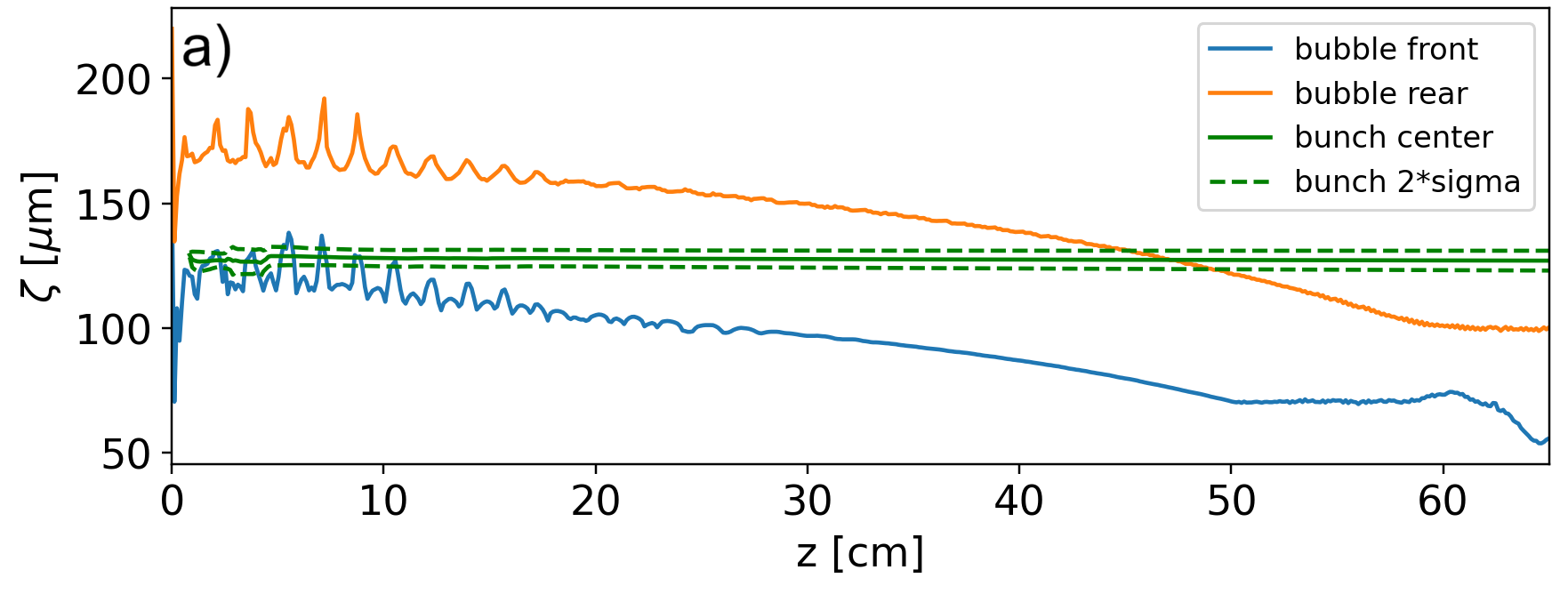}
    \end{subfigure}
    \begin{subfigure}{\columnwidth}
        \centering
        \includegraphics[width=0.99\columnwidth]{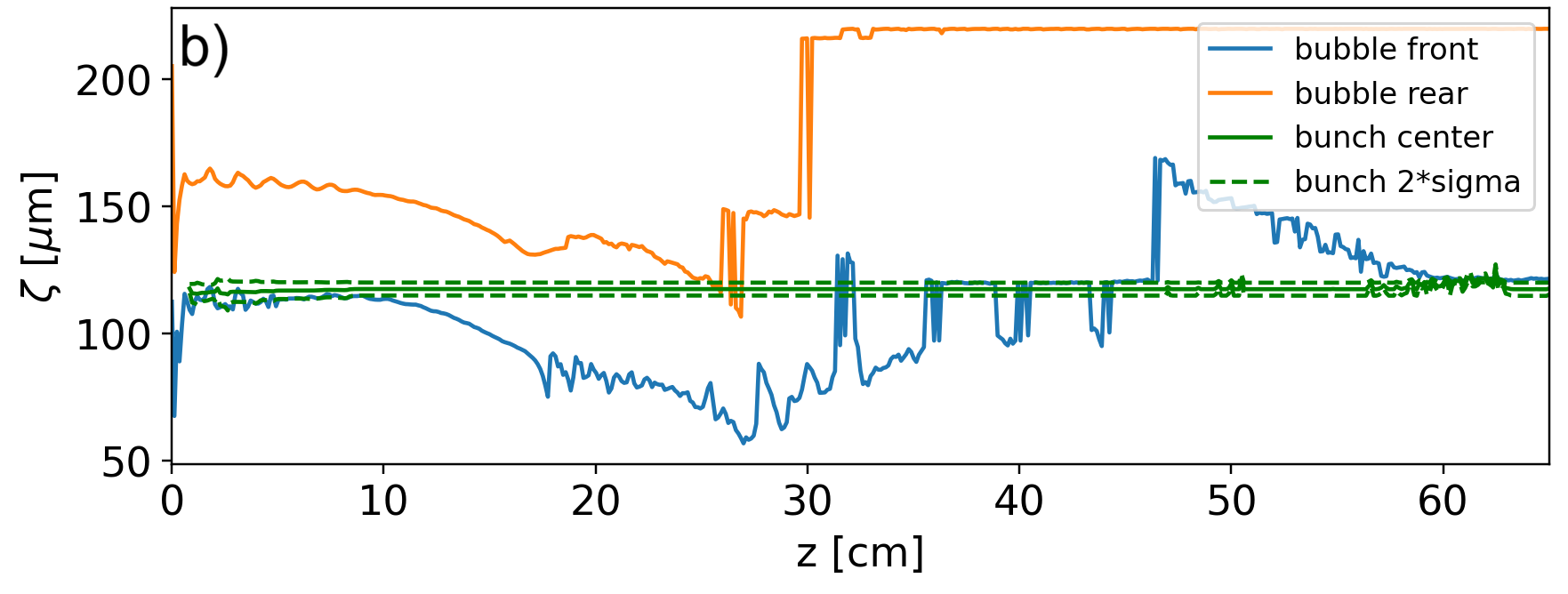}
    \end{subfigure}
    \begin{subfigure}{\columnwidth}
        \centering
        \includegraphics[width=0.99\columnwidth]{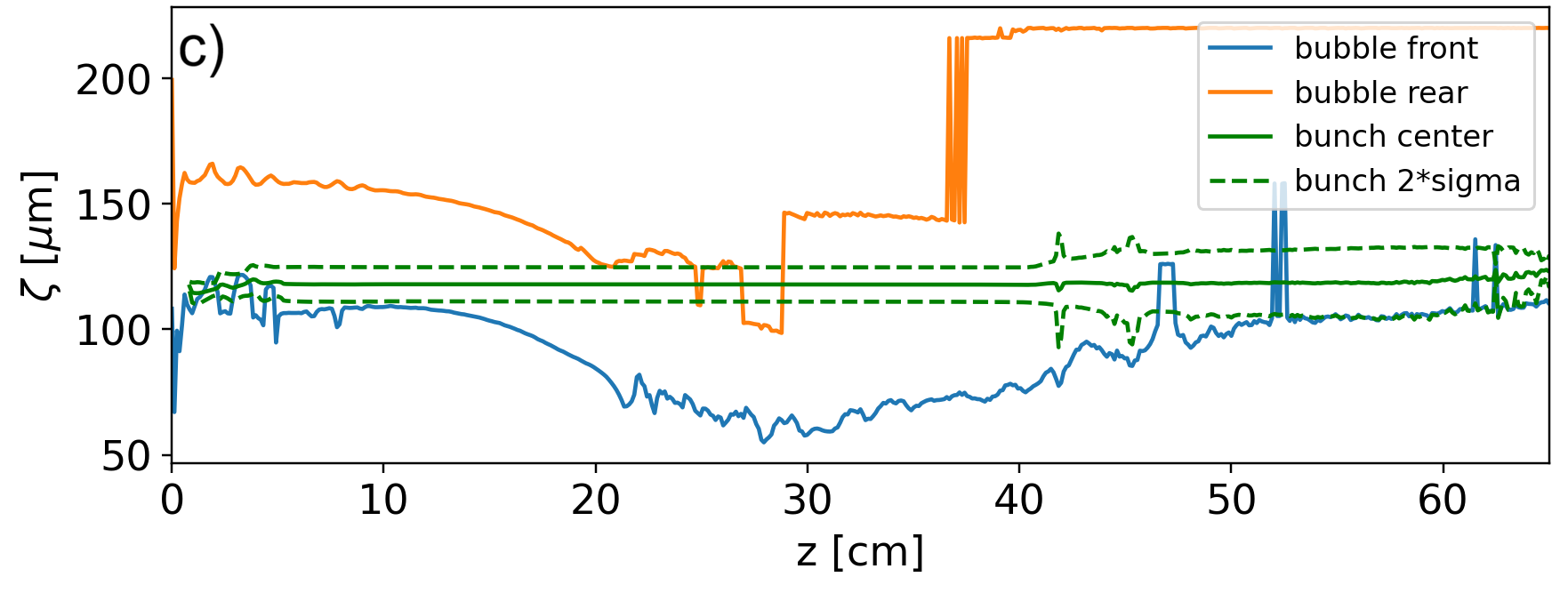}
    \end{subfigure}
\caption{\label{fig:bub} Visualization of the electron bunch co-moving coordinate $\zeta$ (green solid) relative to propagation coordinate $z$ and $2\sigma$ distance about the center relative to the front of the bubble (blue) and the rear of the bubble (orange) for a) the Gaussian, b) HE, and c) HQ case. }
\end{figure}

There are several physical reasons why these pulse shapes perform better than the Gaussian pulse. One is that the spectral content of the laser reduces beating in the laser intensity as it propagates through the plasma,\cite{leemans2014,djordjevic2018} resulting in a more stable accelerating electric field. This can be seen in Fig. \ref{fig:a0_osc}.a). It is noticeable that spectrally shaped pulses do not oscillate significantly relative to the Gaussian laser pulse.\cite{benedetti2012,benedetti2015} Likewise, there is an appreciable increase in the laser intensity over time, particularly for the HQ case. This means that $E_z$ will increase accordingly, further enhancing the acceleration of the electron bunch and likely compensating for any beamloading due to the high bunch charge ($\gtrsim$ 500 pC). This is illustrated by plots of $E_m$ and $Q$ with respect to time as shown in Fig. \ref{fig:a0_osc}.b), with significant charge injection (several hundred pC), in all three cases. Unlike the HE and HQ cases, the Gaussian loses much of the charge during the initial beating of the laser mode. If we focus earlier in time in Fig. \ref{fig:a0_osc}.c), we observe that not only is the Gaussian pulse losing charge over time, with a precipitous drop around 4.5 cm, but the HQ is increasing in the amount of charge injected despite periodic losses due to beating given the extended dopant region (vertical line). The HE case injects less charge overall than the Gaussian or HQ case but maintains it throughout until depletion occurs. We can further visualize the electron bunch relative to the bubble within which it is being accelerated, as seen in Fig. \ref{fig:bub}. Here, the bubble is defined simply as the first bucket in the wake where the electric field is negative. This illustrates how the bunch leaves and reenters the bubble during oscillations correlates with the beating in laser intensity $a_0$.

\begin{figure*}[h!t]
    \centering
    \begin{subfigure}{0.32\textwidth}
        \centering
        \includegraphics[height=4.2cm]{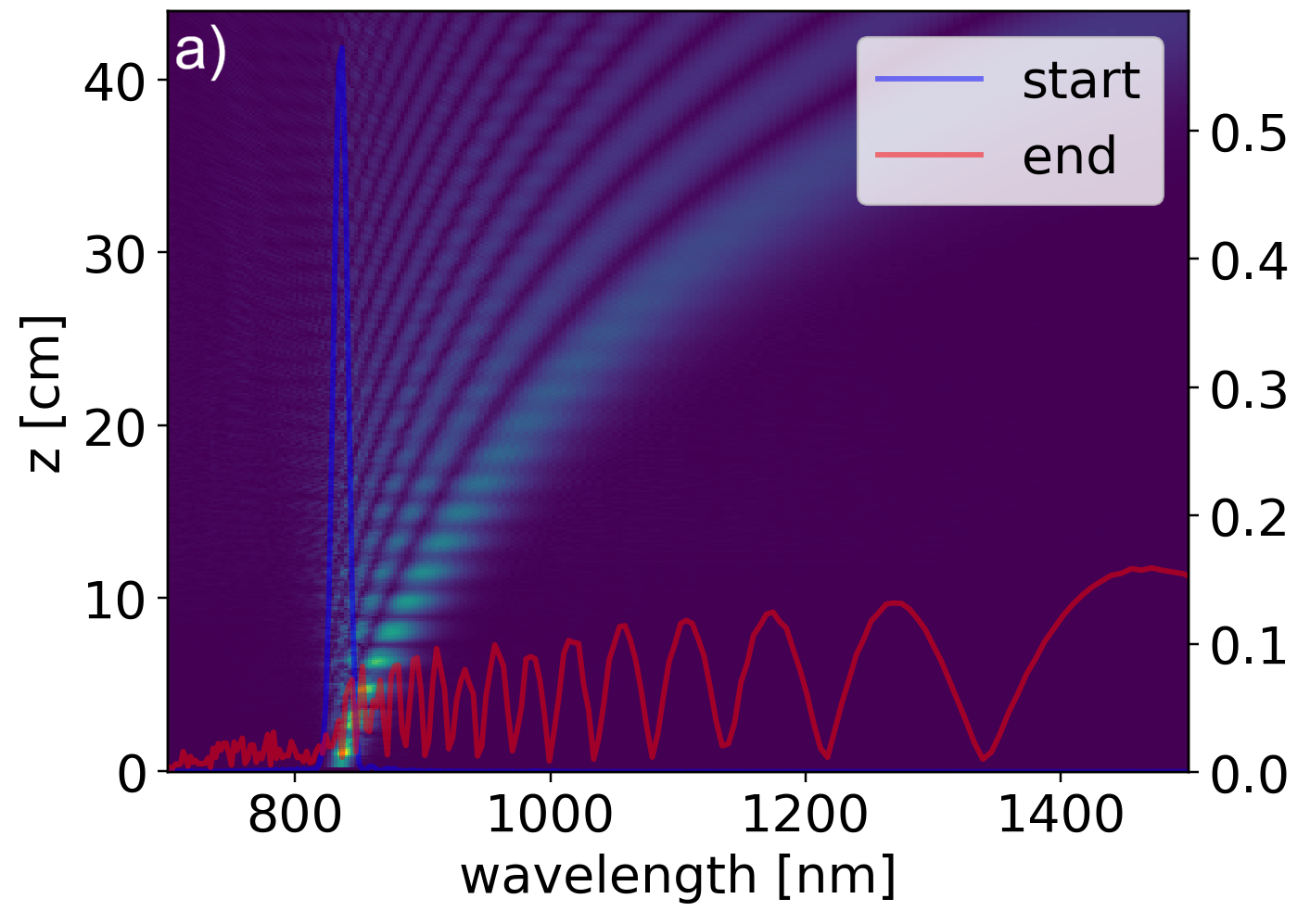}
    \end{subfigure}%
    \begin{subfigure}{0.32\textwidth}
        \centering
        \includegraphics[height=4.2cm]{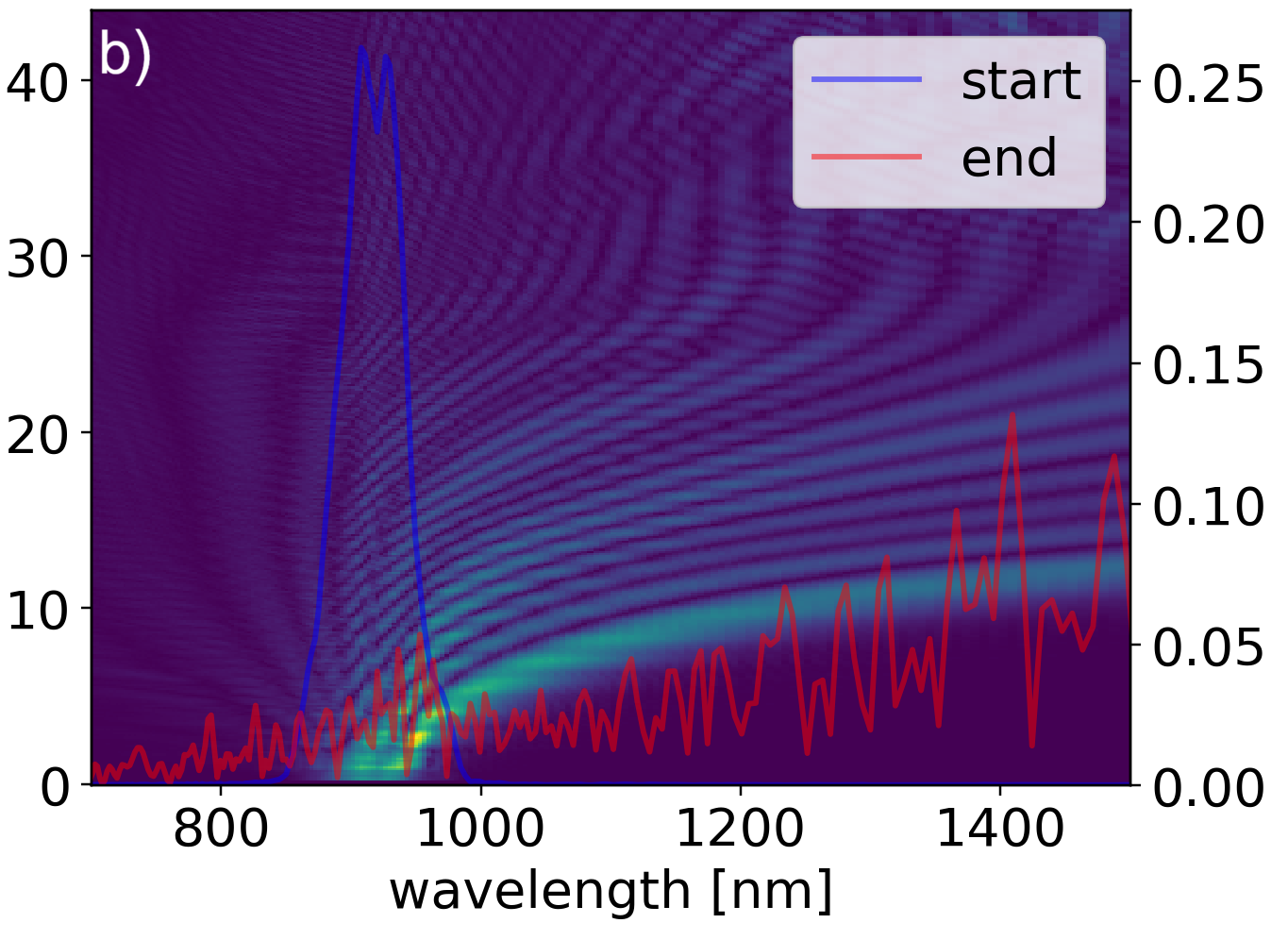}
    \end{subfigure}%
    \begin{subfigure}{0.32\textwidth}
        \centering
        \includegraphics[height=4.2cm]{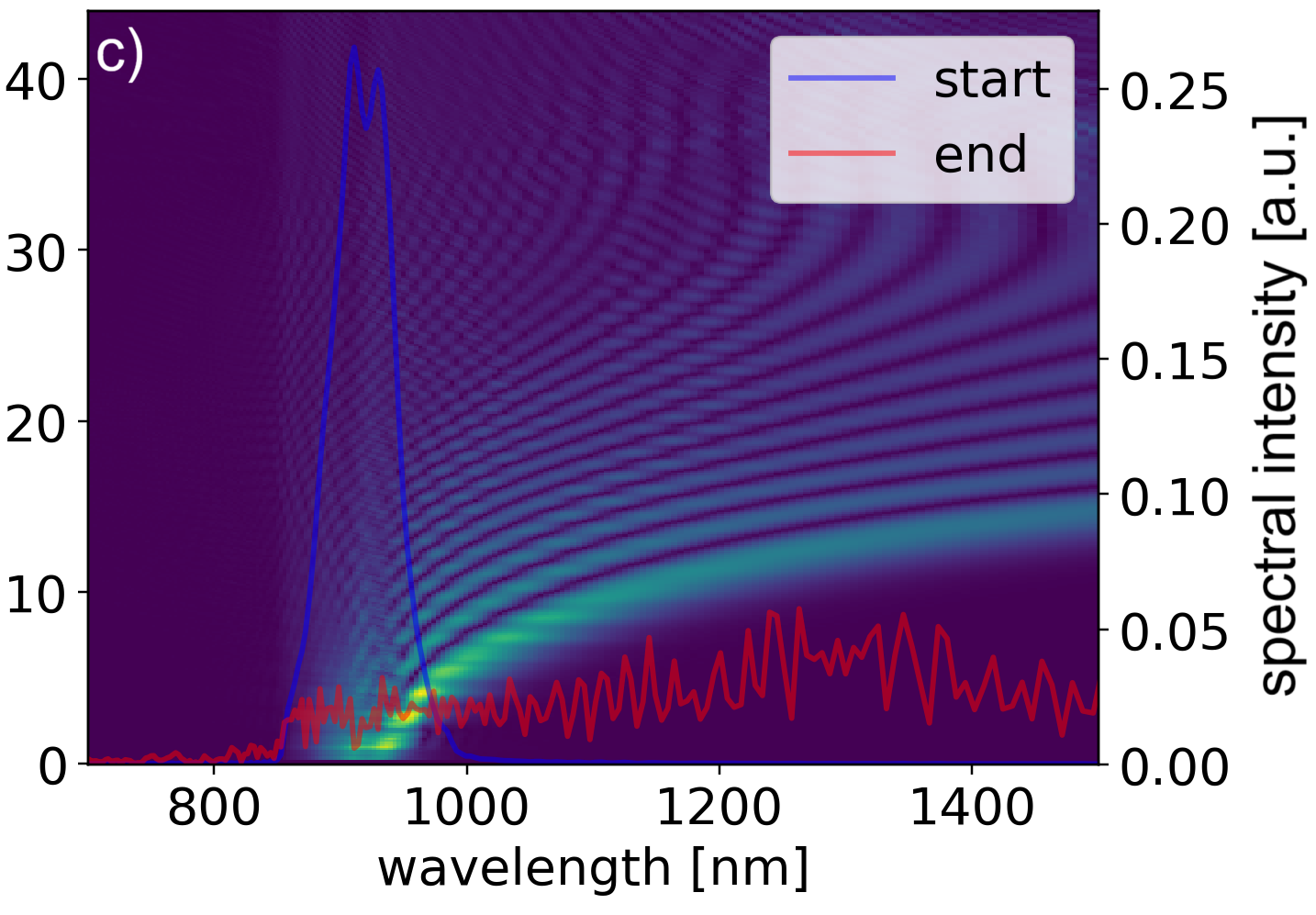}
    \end{subfigure}%

    \caption{\label{fig:lspectra} The laser spectra as a function of time and space for the a) Gaussian, b) HE, and c) HQ cases. The initial spectrum (blue) significantly redshifts as the laser propagates through the plasma (red). 
    }
\end{figure*}

A noticeable difference between the Gaussian pulse and the HE and HQ cases is the significantly more laser redshifting for BO-based cases, as seen in Fig. \ref{fig:lspectra}. The primary reason for this is the density slope in the longitudinal profile, resulting in higher densities as the laser propagates. The characteristic length scale for laser pump depletion\cite{benedetti2015} is $L_{pd} \approx \frac{\omega_0^2}{\omega_p^2} \frac{c\tau_L}{a_0^2} = \frac{n_c}{n_e} \frac{c\tau_L}{a_0^2}$, and the corresponding fractional frequency downshift scales as $\frac{\Delta\omega}{\omega_0} \sim \frac{L}{L_{pd}} \propto n_e$. Likewise, there is self-phase modulation via the plasma index $\eta=\sqrt{1-\omega_p^2/\omega^2}$, which will drive a stronger, typically negative chirp, i.e., more redshift. Lastly, given that there is a plasma gradient, for a slowly varying plasma with dispersion $\omega^2 = \omega_p^2(z,t)+k^2c^2$, the laser frequency evolves roughly as 
\begin{equation}
    \frac{d\omega}{dt} \simeq - \frac{1}{2\omega}\frac{d\omega_p^2}{dt} \approx -\frac{v_g}{2\omega}\frac{d\omega_p^2}{dz},
\end{equation}
where $d/dt \approx v_g d/dz$. So when we have an up-ramp $(dn_e/d_z >0) \rightarrow d\omega/dt <0 $, we have redshifting, while a down-ramp $(dn_e/d_z <0) \rightarrow d\omega/dt >0 $ gives blueshifting. 

\section{Conclusions}
\label{conc}

In this work, we used Bayesian optimization to investigate the effects of spectral shaping and plasma channel parameters on key LWFA figures of merit, i.e., the mean energy of the accelerated electron bunch $E_m$, charge $Q$ and energy spread $\sigma_E$. Spectral shaping alone can have an effect on LWFA, particularly GDD and TOD, greatly increasing $Q$ but not appreciably increasing the bunch energy $E_m$. Incorporating additional laser and plasma channel parameters, such as injection distance, dopant length, dopant fraction, and a linear longitudinal density slope, allowed for significant improvement. The best performer achieved 15 GeV energies within the current parameter boundaries without sacrificing high charge. In fact, not only can we increase energy with a noticeable increase of charge (from 10 to ~500 pC), but we can also achieve comparable energies but with approaching nC charge content levels while keeping laser energy constant. 

It is important to note that experiments involve additional considerations in terms of the final pulse shape beyond just spectral shaping, as performed by the DAZZLER. Most notably, the amplification process that is key to PW laser systems will nonlinearly distort the laser pulse away from the pulse shape chosen by the DAZZLER, since it is typically early on in the laser chain, near the initial oscillator. That means that the spectral phase components, chosen based on the analysis, provided in this study do not exactly correlate to those seen in an experiment. Likewise, not only does each laser have its own distinct spectrum, but the laser spectrum may vary from shot to shot, further complicating effective modeling. Even measuring a pick-off beam for each shot is not fully reliable, as the diagnostic may not be perfectly aligned with the pick-off beam for each shot. However, the plasma is agnostic to how the laser is made, and therefore pulse shapes found in such studies are not rendered irrelevant, just that different laser input parameters may be needed to achieve the same pulse shape. 

In the context of the laser wakefield acceleration, spectral shaping can play a critical role in tailoring the drive laser for optimal wakefield formation and charge injection. By modulating the spectral phase on the basis of numerical optimization, one can achieve high-charge bunches with improved energy spread and beam quality. When combined with advanced plasma channel designs such as tapered HOFI structures, the synergy between shaped pulses and matched guiding profiles offers a promising path to achieving reproducible, high-performance acceleration.

This is a relatively ad hoc application of BO to LWFA and can be expanded further. A more realistic laser modeling approach could be incorporated, i.e., modeling the effects of amplification, compression, focusing, deformable mirrors, etc. Similarly, even more realistic plasma channels could be modeled. For example, a linear density slope was applied across the entire plasma channel in this study, while a realistic, ionized gas jet would have different channel transverse curvatures and profiles based on the local density longitudinally. In addition, multi-fidelity BO can be used,\cite{gratiet2013} and recent work has demonstrated fully multi-objective and multi-fidelity BO for LWFA using hypervolume-based and trust-aware acquisition policies that reduce the cost of exploring Pareto-optimal beam settings,\cite{irshad2023} an approach that could be extended to spectral shaping.

\begin{acknowledgments}
This work was supported by the Defense
Advanced Research Projects Agency (DARPA),  the Director, Office of Science, Office of High Energy Physics, of the U.S. Department of Energy under Contract Nos. DE-AC02-05CH11231 and DE-AC52-07NA27344, and used the computational facilities at the National Energy Research Scientific Computing Center (NERSC) and Livermore Computing. This document has been approved for distribution per Release No. LLNL-JRNL-2012961-DRAFT. This research used the open-source particle-in-cell code WarpX \url{https://github.com/BLAST-WarpX/warpx}. Primary WarpX contributors are with LBNL, LLNL, CEA-LIDYL, SLAC, DESY, CERN, Helion Energy, and TAE Technologies. We acknowledge all WarpX contributors. We would also like to acknowledge Axel Huebl, Joshua Ludwig, and Zoran Djordjevi\'{c} for their advice and help. Manuscript has been approved for distribution per release number LLNL-JRNL--DRAFT. This document was prepared as an account of work sponsored by an agency of the United States government. Neither the United States government nor Lawrence Livermore National Security, LLC, nor any of their employees makes any warranty, expressed or implied, or assumes any legal liability or responsibility for the accuracy, completeness, or usefulness of any information, apparatus, product, or process disclosed, or represents that its use would not infringe privately owned rights. Reference herein to any specific commercial product, process, or service by trade name, trademark, manufacturer, or otherwise does not necessarily constitute or imply its endorsement, recommendation, or favoring by the United States government or Lawrence Livermore National Security, LLC. The views and opinions of authors expressed herein do not necessarily state or reflect those of the United States government or Lawrence Livermore National Security, LLC, and shall not be used for advertising or product endorsement purposes. 
\end{acknowledgments}


\nocite{*}
\bibliography{main}

@PREAMBLE{
 "\providecommand{\noopsort}[1]{}" 
 # "\providecommand{\singleletter}[1]{#1}%" 
}

@article{dann2019,
  title = {Laser wakefield acceleration with active feedback at 5 Hz},
  author = {Dann, S. J. D. and Baird, C. D. and Bourgeois, N. and Chekhlov, O. and Eardley, S. and Gregory, C. D. and Gruse, J.-N. and Hah, J. and Hazra, D. and Hawkes, S. J. and Hooker, C. J. and Krushelnick, K. and Mangles, S. P. D. and Marshall, V. A. and Murphy, C. D. and Najmudin, Z. and Nees, J. A. and Osterhoff, J. and Parry, B. and Pourmoussavi, P. and Rahul, S. V. and Rajeev, P. P. and Rozario, S. and Scott, J. D. E. and Smith, R. A. and Springate, E. and Tang, Y. and Tata, S. and Thomas, A. G. R. and Thornton, C. and Symes, D. R. and Streeter, M. J. V.},
  journal = {Phys. Rev. Accel. Beams},
  volume = {22},
  issue = {4},
  pages = {041303},
  numpages = {12},
  year = {2019},
  month = {Apr},
  publisher = {American Physical Society},
  doi = {10.1103/PhysRevAccelBeams.22.041303},
  url = {https://link.aps.org/doi/10.1103/PhysRevAccelBeams.22.041303}
}

@article{leemans2014,
  title = {Multi-GeV Electron Beams from Capillary-Discharge-Guided Subpetawatt Laser Pulses in the Self-Trapping Regime},
  author = {Leemans, W. P. and Gonsalves, A. J. and Mao, H.-S. and Nakamura, K. and Benedetti, C. and Schroeder, C. B. and T\'oth, Cs. and Daniels, J. and Mittelberger, D. E. and Bulanov, S. S. and Vay, J.-L. and Geddes, C. G. R. and Esarey, E.},
  journal = {Phys. Rev. Lett.},
  volume = {113},
  issue = {24},
  pages = {245002},
  numpages = {5},
  year = {2014},
  month = {Dec},
  publisher = {American Physical Society},
  doi = {10.1103/PhysRevLett.113.245002},
  url = {https://link.aps.org/doi/10.1103/PhysRevLett.113.245002}
}

@article{djordjevic2018,
    author = {Djordjević, B. Z. and Benedetti, C. and Schroeder, C. B. and Esarey, E. and Leemans, W. P.},
    title = {Filtering higher-order laser modes using leaky plasma channels},
    journal = {Physics of Plasmas},
    volume = {25},
    number = {1},
    pages = {013103},
    year = {2018},
    month = {01},
    issn = {1070-664X},
    doi = {10.1063/1.5006198},
    url = {https://doi.org/10.1063/1.5006198},
}

@article{rittershofer2010,
    author = {Rittershofer, W. and Schroeder, C. B. and Esarey, E. and Grüner, F. J. and Leemans, W. P.},
    title = {Tapered plasma channels to phase-lock accelerating and focusing forces in laser-plasma accelerators},
    journal = {Physics of Plasmas},
    volume = {17},
    number = {6},
    pages = {063104},
    year = {2010},
    month = {06},
    issn = {1070-664X},
    doi = {10.1063/1.3430638},
    url = {https://doi.org/10.1063/1.3430638},
}

@article{djordjevic2019,
    author = {Djordjević, B. Z. and Benedetti, C. and Schroeder, C. B. and Esarey, E.},
    title = {Chromatic matching in a plasma undulator},
    journal = {Physics of Plasmas},
    volume = {26},
    number = {11},
    pages = {113102},
    year = {2019},
    month = {11},
    issn = {1070-664X},
    doi = {10.1063/1.5120868},
    url = {https://doi.org/10.1063/1.5120868},
}

@article{aniculaesei2023,
    author = {Aniculaesei, Constantin and Ha, Thanh and Yoffe, Samuel and Labun, Lance and Milton, Stephen and McCary, Edward and Spinks, Michael M. and Quevedo, Hernan J. and Labun, Ou Z. and Sain, Ritwik and Hannasch, Andrea and Zgadzaj, Rafal and Pagano, Isabella and Franco-Altamirano, Jose A. and Ringuette, Martin L. and Gaul, Erhart and Luedtke, Scott V. and Tiwari, Ganesh and Ersfeld, Bernhard and Brunetti, Enrico and Ruhl, Hartmut and Ditmire, Todd and Bruce, Sandra and Donovan, Michael E. and Downer, Michael C. and Jaroszynski, Dino A. and Hegelich, Bjorn Manuel},
    title = {The acceleration of a high-charge electron bunch to 10 GeV in a 10-cm nanoparticle-assisted wakefield accelerator},
    journal = {Matter and Radiation at Extremes},
    volume = {9},
    number = {1},
    pages = {014001},
    year = {2023},
    month = {11},
    issn = {2468-2047},
    doi = {10.1063/5.0161687},
    url = {https://doi.org/10.1063/5.0161687},
}

@article{streeter2018,
    author = {Streeter, M. J. V. and Dann, S. J. D. and Scott, J. D. E. and Baird, C. D. and Murphy, C. D. and Eardley, S. and Smith, R. A. and Rozario, S. and Gruse, J.-N. and Mangles, S. P. D. and Najmudin, Z. and Tata, S. and Krishnamurthy, M. and Rahul, S. V. and Hazra, D. and Pourmoussavi, P. and Osterhoff, J. and Hah, J. and Bourgeois, N. and Thornton, C. and Gregory, C. D. and Hooker, C. J. and Chekhlov, O. and Hawkes, S. J. and Parry, B. and Marshall, V. A. and Tang, Y. and Springate, E. and Rajeev, P. P. and Thomas, A. G. R. and Symes, D. R.},
    title = {Temporal feedback control of high-intensity laser pulses to optimize ultrafast heating of atomic clusters},
    journal = {Applied Physics Letters},
    volume = {112},
    number = {24},
    pages = {244101},
    year = {2018},
    month = {06},
    issn = {0003-6951},
    doi = {10.1063/1.5027297},
    url = {https://doi.org/10.1063/1.5027297},
}

@article{backhouse2025,
  title = {Spectral phase control for optimized ionization injection in laser wakefield acceleration},
  author = {Backhouse, M. P. and Dickson, L. T. and Moulanier, I. and Massimo, F. and Cobo, C. C. and Filippi, F. and Gustafsson, C. and Lofquist, E. and Svendsen, K. and Streeter, M. J. V. and Shalloo, R. J. and Vasilovici, O. and Ballage, C. and Dann, S. J. D. and Murphy, C. D. and Najmudin, Z. and Dobosz Dufr\'enoy, S. and Lundh, O. and Cros, B.},
  journal = {Phys. Rev. Accel. Beams},
  volume = {28},
  issue = {12},
  pages = {121301},
  numpages = {6},
  year = {2025},
  month = {Dec},
  publisher = {American Physical Society},
  doi = {10.1103/9gjq-htzn},
  url = {https://link.aps.org/doi/10.1103/9gjq-htzn}
}

@article{ludwig2025,
  author    = {Ludwig, J. D. and Wilks, S. C. and Kemp, A. J. and Williams, G. J. and 
               Lemos, N. and Rockafellow, E. and Miao, B. and Shrock, J. E. and 
               Milchberg, H. M. and Vay, J.-L. and Huebl, A. and Lehe, R. and 
               Cimmino, A. and Versaci, R. and Bulanov, S. V. and Valenta, P. and 
               Tang, V. and Reagan, B. A.},
  title     = {Laser based 100 GeV electron acceleration scheme for muon production},
  journal   = {Scientific Reports},
  year      = {2025},
  volume    = {15},
  number    = {1},
  pages     = {25902},
  doi       = {10.1038/s41598-025-95440-w},
  url       = {https://doi.org/10.1038/s41598-025-95440-w}
}

@article{esarey2009,
  author       = {Esarey, E. and Schroeder, C. B. and Leemans, W. P.},
  title        = {{Physics of laser‐driven plasma‐based electron accelerators}},
  journal      = {Reviews of Modern Physics},
  volume       = {81},
  number       = {3},
  pages        = {1229--1285},
  year         = {2009},
  month        = {August},
  doi          = {10.1103/RevModPhys.81.1229}
}

@article{Roussel2024Bayesian,
  author       = {Roussel, Ryan and Edelen, Auralee L. and Boltz, Tobias and Kennedy, Dylan and Zhang, Zhe and Ji, Fuhao and Huang, Xiaobiao and Ratner, Daniel and Santamaria Garcia, Andrea and Xu, Chenran and Kaiser, Jan and Pousa, Angel Ferran and Eichler, Annika and L{\"u}bsen, Jannis O. and Isenberg, Natalie M. and Gao, Yuan and Kuklev, Nikita and Martinez, Jose and Mustapha, Brahim and Kain, Verena and Mayes, Christopher and Lin, Weijian and Liuzzo, Simone Maria and St. John, Jason and Streeter, Matthew J.V. and Lehe, Remi and Neiswanger, Willie},
  title        = {Bayesian Optimization Algorithms for Accelerator Physics},
  journal      = {Physical Review Accelerators and Beams},
  volume       = {27},
  number       = {8},
  pages        = {084801},
  year         = {2024},
  doi          = {10.1103/PhysRevAccelBeams.27.084801}
}

@article{saves2024smt,
        author = {P. Saves and R. Lafage and N. Bartoli and Y. Diouane and J. Bussemaker and T. Lefebvre and J. T. Hwang and J. Morlier and J. R. R. A. Martins},
        title = {{SMT 2.0: A} Surrogate Modeling Toolbox with a focus on Hierarchical and Mixed Variables Gaussian Processes},
        journal = {Advances in Engineering Sofware},
        year = {2024},
        volume = {188},
        pages = {103571},
        doi = {https://doi.org/10.1016/j.advengsoft.2023.103571}}

@article{tajima1979,
  author       = {Tajima, T. and Dawson, J. M.},
  title        = {{Laser Electron Accelerator}},
  journal      = {Physical Review Letters},
  volume       = {43},
  number       = {4},
  pages        = {267--270},
  year         = {1979},
  doi          = {10.1103/PhysRevLett.43.267}
}

@article{cole2015,
  author       = {Cole, J. M. and Wood, J. C. and Lopes, N. C. and Poder, K. and Abel, R. L. and Alatabi, S. and Bryant, J. S. J. and others},
  title        = {{Laser-wakefield accelerators as hard x-ray sources for 3D medical imaging of human bone}},
  journal      = {Scientific Reports},
  volume       = {5},
  pages        = {13244},
  year         = {2015},
  doi          = {10.1038/srep13244}
}

@article{schroeder2010,
  author       = {Schroeder, C. B. and Esarey, E. and Geddes, C. G. R. and Benedetti, C. and Leemans, W. P.},
  title        = {{Physics considerations for laser‐plasma linear colliders}},
  journal      = {Physical Review Special Topics – Accelerators and Beams},
  volume       = {13},
  pages        = {101301},
  year         = {2010},
  doi          = {10.1103/PhysRevSTAB.13.101301}
}

@article{schroeder2016,
  author       = {Schroeder, C. B. and others},
  title        = {{A linear electron–positron collider based on laser-plasma accelerators using hollow plasma channels}},
  journal      = {Nuclear Instruments and Methods in Physics Research Section A},
  volume       = {829},
  pages        = {32--37},
  year         = {2016},
  doi          = {10.1016/j.nima.2016.06.009}  
}

@article{akhiezer1956,
  author       = {A. I. Akhiezer and R. V. Polovin},
  title        = {{Theory of Wave Motion of an Electron Plasma}},
  journal      = {Soviet Physics JETP},
  volume       = {3},
  number       = {5},
  pages        = {696--705},
  year         = {1956},
  note         = {Received: March 19, 1955; in Russian},
}

@inproceedings{albert2024,
  author       = {Albert, F{\’e}licie},
  title        = {{X-ray Light Sources driven by laser-plasma acceleration for high energy density science applications}},
  booktitle    = {High-Brightness Sources and Light-Driven Interactions Congress},
  series       = {Technical Digest Series (Optica Publishing Group)},
  year         = {2024},
  paper        = {ETu3A.1},
}

@article{geddes2004,
  author       = {Geddes, C. G. R. and Tóth, Cs. and van Tilborg, J. and Esarey, E. and Schroeder, C. B. and Bruhwiler, D. and Nieter, C. and Cary, J. and Leemans, W. P.},
  title        = {{High-quality electron beams from a laser wakefield accelerator using plasma-channel guiding}},
  journal      = {Nature},
  volume       = {431},
  number       = {7008},
  pages        = {538--541},
  year         = {2004},
  doi          = {10.1038/nature02900}
}

@article{leemans2006,
  author       = {Leemans, W. P. and Nagler, B. and Gonsalves, A. J. and Tóth, Cs. and Nakamura, K. and Geddes, C. G. R. and Esarey, E. and Schroeder, C. B. and Hooker, S. M.},
  title        = {{GeV electron beams from a centimetre-scale accelerator}},
  journal      = {Nature Physics},
  volume       = {2},
  number       = {10},
  pages        = {696--699},
  year         = {2006},
  doi          = {10.1038/nphys418}
}

@article{gonsalves2019,
  author       = {Gonsalves, A. J. and Nakamura, K. and Daniels, J. and Benedetti, C. and Pieronek, C. and de Raadt, T. C. H. and Steinke, S. and Bin, J. H. and Bulanov, S. S. and van Tilborg, J. and Geddes, C. G. R. and Schroeder, C. B. and Tóth, Cs. and Esarey, E. and Swanson, K. K. and Fan-Chen, G. and Bagdasarov, G. and Bobrova, N. and Gasilov, V. and Korn, G. and Sasorov, P. and Leemans, W. P.},
  title        = {{Petawatt laser guiding and electron beam acceleration to 8 GeV in a laser-heated capillary discharge waveguide}},
  journal      = {Physical Review Letters},
  volume       = {122},
  pages        = {084801},
  year         = {2019},
  doi          = {10.1103/PhysRevLett.122.084801}
}

@article{barber2025,
  author       = {Barber, S. K. and Kohrell, F. and Doss, C. E. and Jensen, K. and Berger, C. and Isono, F. and Eisentraut, Z. and Schröder, S. and Gonsalves, A. J. and Nakamura, K. and Plateau, G. R. and van Mourik, R. A. and Gracia-Linares, M. and Labun, L. and Hegelich, B. M. and Milton, S. V. and Geddes, C. G. R. and Osterhoff, J. and Schroeder, C. B. and Esarey, E. H. and van Tilborg, J.},
  title        = {{Greater than 1000-fold gain in a free-electron laser driven by a laser-plasma accelerator with high reliability}},
  journal      = {Physical Review Letters},
  volume       = {135},
  pages        = {055001},
  year         = {2025},
  doi          = {10.1103/PhysRevLett.135.055001}
}

@article{durfee1993,
  author       = {Durfee, C. G. and Milchberg, H. M.},
  title        = {{Light pipe for high intensity laser pulses}},
  journal      = {Physical Review Letters},
  volume       = {71},
  pages        = {2409--2412},
  year         = {1993},
  doi          = {10.1103/PhysRevLett.71.2409}
}

@article{shalloo2020,
  author       = {Shalloo, R. J. and Arran, C. and Corner, L. and Cheung, G. and Picksley, A. and Holloway, J. and Thornton, C. and Wing, M. and Hook, S. and Walczak, R. and others},
  title        = {{Hydrodynamic optical-field-ionized plasma channels}},
  journal      = {Physical Review E},
  volume       = {101},
  number       = {5},
  pages        = {053201},
  year         = {2020},
  doi          = {10.1103/PhysRevE.101.053201}
}

@article{picksley2024,
  author       = {Picksley, A. and Stackhouse, J. and Benedetti, C. and Nakamura, K. and Tsai, H. E. and Li, R. and Miao, B. and Shrock, J. E. and Rockafellow, E. and Milchberg, H. M. and Schroeder, C. B. and van Tilborg, J. and Esarey, E. and Geddes, C. G. R. and Gonsalves, A. J.},
  title        = {{Matched Guiding and Controlled Injection in Dark-Current-Free, 10-GeV-Class, Channel-Guided Laser-Plasma Accelerators}},
  journal      = {Physical Review Letters},
  volume       = {133},
  pages        = {255001},
  year         = {2024},
  doi          = {10.1103/PhysRevLett.133.255001}
}

@article{frazier2018,
  author       = {Frazier, Peter I.},
  title        = {{A Tutorial on Bayesian Optimization}},
  journal      = {arXiv},
  volume       = {abs/1807.02811},
  pages        = {–},  
  year         = {2018},
  doi          = {10.48550/arXiv.1807.02811}
}

@article{jalas2021,
  author       = {Jalas, Sören and Kirchen, Manuel and Messner, Philipp and Winkler, Paul and Hübner, Lars and Dirkwinkel, Julian and Schnepp, Matthias and Lehe, Rémi and Maier, Andreas R.},
  title        = {{Bayesian Optimization of a Laser-Plasma Accelerator}},
  journal      = {Physical Review Letters},
  volume       = {126},
  number       = {10},
  pages        = {104801},
  year         = {2021},
  doi          = {10.1103/PhysRevLett.126.104801}
}

@article{loughran2023,
  author       = {Loughran, B. and Streeter, M. J. V. and Ahmed, H. and Astbury, S. and Balcazar, M. and Borghesi, M. and Bourgeois, N. and Curry, C. B. and Dann, S. J. D. and DiIorio, S. and Dover, N. P. and Dzelzanis, T. and Ettlinger, O. C. and Gauthier, M. and Giuffrida, L. and Glenn, G. D. and Glenzer, S. H. and Green, J. S. and Gray, R. J. and Hyland, C. and Istokskaia, V. and King, M. and Margarone, D. and McCusker, O. and McKenna, P. and Najmudin, Z. and Parisuaña, C. and Parsons, P. and Spindloe, C. and Symes, D. R. and Thomas, A. G. R. and Treffert, F. and Xu, N. and Palmer, C. A. J.},
  title        = {{Automated Control and Optimization of Laser-Driven Ion Acceleration}},
  journal      = {High Power Laser Science and Engineering},
  volume       = {11},
  pages        = {e35},
  year         = {2023},
  doi          = {10.1017/hpl.2023.23}
}

@article{ziegler2021,
  author       = {Ziegler, Thomas and others},
  title        = {{Proton beam quality enhancement by spectral phase modulation in high-power laser driven ion acceleration experiments}},
  journal      = {Scientific Reports},
  volume       = {11},
  pages        = {3536},
  year         = {2021},
  doi          = {10.1038/s41598-021-86547-x}
}

@book{trebino2000,
  author       = {Trebino, Rick},
  title        = {Frequency-Resolved Optical Gating: The Measurement of Ultrashort Laser Pulses},
  publisher    = {Springer},
  address      = {New York, NY},
  year         = {2000},
  isbn         = {978-1-4020-7066-2},
  doi          = {10.1007/978-1-4615-1181-6}
}

@article{cook2025a,
  author       = {Cook, Nathan M. and Wolfinger, Kathryn and others},
  title        = {{Hydrodynamic modeling of plasma channel systems for laser plasma accelerators}},
  journal      = {Physics of Plasmas},
  year         = {2025},
  note         = {submitted},  
  doi          = {10.1063/…}  
}

@book{agrawal2013,
  author    = {Agrawal, Govind P.},
  title     = {Nonlinear Fiber Optics},
  edition   = {5},
  publisher = {Academic Press},
  year      = {2013},
  isbn      = {978-0123970237}
}

@inproceedings{cook2020,
  author       = {Cook, Nathan M. and Carlsson, J. and Moeller, P. and Nagler, R. and Tzeferacos, P.},
  title        = {{Modeling of capillary discharge plasmas for wakefield acceleration and beam transport}},
  booktitle    = {J. Phys.: Conf. Ser.},
  volume       = {1596},
  pages        = {012063},
  year         = {2020}
}

@article{diaw2022,
  author       = {Diaw, Abdourahmane and Coleman, S. and Cook, N. M. and Edelen, J. and Hansen, E. C. and Tzeferacos, P.},
  title        = {{Impact of Electron Transport Models on Capillary Discharge Plasmas}},
  journal      = {Physics of Plasmas},
  volume       = {29},
  number       = {6},
  pages        = {063101},
  year         = {2022},
  doi          = {10.1063/5.0091809}
}

@inproceedings{cook2024,
  author       = {Cook, Nathan M. and Hall, C. and Wolfinger, K. and Coleman, S. and Picksley, A. and Gonsalves, A. J. and Schroeder, C. B. and Benedetti, C.},
  title        = {{Design and modeling of HOFI plasma channels for laser plasma accelerators}},
  booktitle    = {IPAC '24 Proceedings},
  year         = {2024},
  doi          = {10.18429/JACoW-IPAC2024-MOPR56}
}

@article{kim2017,
  author       = {Kim, H. T. and Pae, K. H. and Cha, H. J. and Kim, I. J. and Yu, T. J. and Sung, J. H. and Lee, S. K. and Jeong, T. M. and Lee, J.},
  title        = {{Stable multi‐GeV electron accelerator driven by waveform‐controlled PW laser pulses}},
  journal      = {Scientific Reports},
  volume       = {7},
  pages        = {10203},
  year         = {2017},
  doi          = {10.1038/s41598-017-09267-1}
}

@article{valenta2025,
  title = {Bayesian optimization of electron energy from laser wakefield accelerators},
  author = {Valenta, P. and Esirkepov, T. Zh. and Ludwig, J. D. and Wilks, S. C. and Bulanov, S. V.},
  journal = {Phys. Rev. Accel. Beams},
  volume = {28},
  issue = {9},
  pages = {094601},
  numpages = {14},
  year = {2025},
  month = {Sep},
  publisher = {American Physical Society},
  doi = {10.1103/knh7-hbr3},
  url = {https://link.aps.org/doi/10.1103/knh7-hbr3}
}

@article{grigoriadis2023,
  author       = {Grigoriadis, S. and Streeter, M. J. V. and Dann, S. J. D. and Symes, D. R. and Najmudin, Z.},
  title        = {{Efficient plasma electron accelerator driven by linearly chirped multi‐10‐TW laser pulses}},
  journal      = {Scientific Reports},
  volume       = {13},
  pages        = {28755},
  year         = {2023},
  doi          = {10.1038/s41598-023-28755-1}
}

@article{pathak2018,
    author = {Pathak, Naveen and Zhidkov, Alexei and Hosokai, Tomonao and Kodama, Ryosuke},
    title = {Spectral effects in the propagation of chirped laser pulses in uniform underdense plasma},
    journal = {Physics of Plasmas},
    volume = {25},
    number = {1},
    pages = {013119},
    year = {2018},
    month = {01},
    issn = {1070-664X},
    doi = {10.1063/1.5011081},
    url = {https://doi.org/10.1063/1.5011081},
}

@article{mariscal2024,
  author       = {Mariscal, D. A. and Djordjevic, B. Z. and Anirudh, R. and Jayaraman-Thiagarajan, J. and Grace, E. S. and Simpson, R. A. and Swanson, K. K. and Galvin, T. C. and Mittelberger, D. and Heebner, J. E. and Muir, R. and Folsom, E. and Hill, M. P. and Feister, S. and Ito, E. and Valdez-Sereno, K. and Rocca, J. J. and Park, J. and Wang, S. and Hollinger, R. and Nedbailo, R. and Sullivan, B. and Zeraouli, G. and Shukla, A. and Turaga, P. and Sarkar, A. and Van Essen, B. and Liu, S. and Spears, B. and Bremer, P. T. and Ma, T.},
  title        = {{Toward machine-learning-assisted PW-class high-repetition-rate experiments with solid targets}},
  journal      = {Physics of Plasmas},
  volume       = {31},
  number       = {7},
  pages        = {073105},
  year         = {2024},
  doi          = {10.1063/5.0190553}
}

@article{lehe2015,
  author       = {Lehe, R. and Kirchen, M. and Andriyash, I. A. and Godfrey, B. B. and Vay, J.-L.},
  title        = {{A spectral, quasi-cylindrical and dispersion-free Particle-In-Cell algorithm}},
  journal      = {Physical Review Accelerators and Beams / arXiv preprint},
  year         = {2015},
  note         = {arXiv:1507.04790}
}

@article{tzoufras2008,
  author       = {Tzoufras, M. and Lu, W. and Tsung, F. S. and Huang, C. and Mori, W. B. and Katsouleas, T. and Vieira, J. and Fonseca, R. A. and Silva, L. O.},
  title        = {{Beam loading in the nonlinear regime of plasma-based acceleration}},
  journal      = {Physical Review Letters},
  volume       = {101},
  pages        = {145002},
  year         = {2008},
  doi          = {10.1103/PhysRevLett.101.145002}
}

@article{Fedeli2022,
  author={Fedeli, Luca and Huebl, Axel and Boillod-Cerneux, France and Clark, Thomas and Gott, Kevin and Hillairet, Conrad and Jaure, Stephan and Leblanc, Adrien and Lehe, Rémi and Myers, Andrew and Piechurski, Christelle and Sato, Mitsuhisa and Zaim, Neïl and Zhang, Weiqun and Vay, Jean-Luc and Vincenti, Henri},
  journal={SC22: International Conference for High Performance Computing, Networking, Storage and Analysis}, 
  title={Pushing the Frontier in the Design of Laser-Based Electron Accelerators with Groundbreaking Mesh-Refined Particle-In-Cell Simulations on Exascale-Class Supercomputers}, 
  year={2022},
  volume={},
  number={},
  pages={1-12},
  doi={10.1109/SC41404.2022.00008}}

@article{lu2006,
  author       = {Lu, W. and Huang, C. and Zhou, M. and Mori, W. B. and Katsouleas, T.},
  title        = {{Nonlinear theory for relativistic plasma wakefields in the blowout regime}},
  journal      = {Physical Review Letters},
  volume       = {96},
  pages        = {165002},
  year         = {2006},
  doi          = {10.1103/PhysRevLett.96.165002}
}

@article{vay2016,
    author = "Vay, J. -L. and Lehe, R. and Vincenti, H. and Godfrey, B. B. and Haber, I. and Lee, P.",
    editor = "Dorda, Ulrich and Assmann, Ralph and Grebenyuk, Julia and Bernhard, Holzer and Massimo, Ferrario and Andrei, Seryi and Arnd, Specka and Alban, Mosnier and Jens, Osterhoff",
    title = "{Recent advances in high-performance modeling of plasma-based acceleration using the full PIC method}",
    doi = "10.1016/j.nima.2015.12.033",
    journal = "Nucl. Instrum. Meth. A",
    volume = "829",
    pages = "353--357",
    year = "2016"
}

@article{miao2020,
  title = {Optical Guiding in Meter-Scale Plasma Waveguides},
  author = {Miao, B. and Feder, L. and Shrock, J. E. and Goffin, A. and Milchberg, H. M.},
  journal = {Phys. Rev. Lett.},
  volume = {125},
  issue = {7},
  pages = {074801},
  numpages = {7},
  year = {2020},
  month = {Aug},
  publisher = {American Physical Society},
  doi = {10.1103/PhysRevLett.125.074801},
  url = {https://link.aps.org/doi/10.1103/PhysRevLett.125.074801}
}

@article{miao2022,
  title = {Multi-GeV Electron Bunches from an All-Optical Laser Wakefield Accelerator},
  author = {Miao, B. and Shrock, J. E. and Feder, L. and Hollinger, R. C. and Morrison, J. and Nedbailo, R. and Picksley, A. and Song, H. and Wang, S. and Rocca, J. J. and Milchberg, H. M.},
  journal = {Phys. Rev. X},
  volume = {12},
  issue = {3},
  pages = {031038},
  numpages = {17},
  year = {2022},
  month = {Sep},
  publisher = {American Physical Society},
  doi = {10.1103/PhysRevX.12.031038},
  url = {https://link.aps.org/doi/10.1103/PhysRevX.12.031038}
}

@article{shapoval2024,
  author       = {Shapoval, O. and Zoni, E. and Lehe, R. and Thévenet, M. and Vay, J.-L.},
  title        = {{Pseudospectral particle-in-cell formulation with arbitrary charge and current-density time dependencies for the modeling of relativistic plasmas}},
  journal      = {Physical Review E},
  volume       = {110},
  pages        = {025206},
  year         = {2024},
  doi          = {10.1103/PhysRevE.110.025206}
}

@article{albert2016,
  author       = {Albert, F. and Thomas, A. G. R.},
  title        = {{Applications of laser wakefield accelerator-based light sources}},
  journal      = {Plasma Physics and Controlled Fusion},
  volume       = {58},
  number       = {10},
  pages        = {103001},
  year         = {2016},
  doi          = {10.1088/0741-3335/58/10/103001}
}

@article{terzani2025,
  title = {Measurement of directional muon beams generated at the Berkeley Lab Laser Accelerator},
  author = {Terzani, Davide and Kisyov, Stanimir and Greenberg, Stephen and Le Pottier, Luc and Mironova, Maria and Picksley, Alex and Stackhouse, Joshua and Tsai, Hai-En and Li, Raymond and Rockafellow, Ela and Miao, Bo and Shrock, Jaron E. and Heim, Timon and Garcia-Sciveres, Maurice and Benedetti, Carlo and Valentine, John and Milchberg, Howard M. and Nakamura, Kei and Gonsalves, Anthony J. and van Tilborg, Jeroen and Schroeder, Carl B. and Esarey, Eric and Geddes, Cameron G. R.},
  journal = {Phys. Rev. Accel. Beams},
  volume = {28},
  number = {10},
  pages = {103401},
  year = {2025},
  doi = {10.1103/kxjr-h7zs},
}

@inproceedings{benedetti2018,
  author    = {Benedetti, Carlo
               and Schroeder, Carl
               and Mehrling, Timon
               and Djordjevic, Blagoje
               and Bulanov, Stepan
               and Geddes, Cameron
               and Esarey, Eric
               and Leemans, Wim},
  title     = {{INF\&RNO} Modeling of 10 GeV-Class Electron Beams from a Laser-Plasma Accelerator Driven by the BELLA Laser},
  booktitle = {2018 IEEE Advanced Accelerator Concepts Workshop (AAC)},
  year      = {2018},
  month     = aug,
  pages     = {1--5},
  publisher = {IEEE},
  address   = {Breckenridge, CO, USA},
}

@article{benedetti2012,
    author = {Benedetti, C. and Schroeder, C. B. and Esarey, E. and Leemans, W. P.},
    title = {Quasi-matched propagation of ultra-short, intense laser pulses in plasma channels},
    journal = {Physics of Plasmas},
    volume = {19},
    number = {5},
    pages = {053101},
    year = {2012},
    month = {05},
    issn = {1070-664X}, 
     doi = {10.1063/1.4707393},
    url = {https://doi.org/10.1063/1.4707393},
}

@article{rockafellow2025,
    author = {Rockafellow, E. and Miao, B. and Shrock, J. E. and Sloss, A. and Le, M. S. and Hancock, S. W. and Zahedpour, S. and Hollinger, R. C. and Wang, S. and King, J. and Zhang, P. and Šišma, J. and Grittani, G. M. and Versaci, R. and Gordon, D. F. and Williams, G. J. and Reagan, B. A. and Rocca, J. J. and Milchberg, H. M.},
    title = {Development of a high charge 10 GeV laser electron accelerator},
    journal = {Physics of Plasmas},
    volume = {32},
    number = {5},
    pages = {053102},
    year = {2025},
    month = {05},
    issn = {1070-664X},
    doi = {10.1063/5.0265640},
    url = {https://doi.org/10.1063/5.0265640},
}

@article{gratiet2013,
author = {Le Gratiet, Loic},
title = {Bayesian Analysis of Hierarchical Multifidelity Codes},
journal = {SIAM/ASA Journal on Uncertainty Quantification},
volume = {1},
number = {1},
pages = {244-269},
year = {2013},
doi = {10.1137/120884122},
URL = { 
        https://doi.org/10.1137/120884122
},
}

@article{irshad2023,
  title = {Multi-objective and multi-fidelity Bayesian optimization of laser-plasma acceleration},
  author = {Irshad, F. and Karsch, S. and D\"opp, A.},
  journal = {Phys. Rev. Res.},
  volume = {5},
  issue = {1},
  pages = {013063},
  numpages = {10},
  year = {2023},
  month = {Jan},
  publisher = {American Physical Society},
  doi = {10.1103/PhysRevResearch.5.013063},
  url = {https://link.aps.org/doi/10.1103/PhysRevResearch.5.013063}
}

@article{benedetti2015,
  title = {Pulse evolution and plasma-wave phase velocity in channel-guided laser-plasma accelerators},
  author = {Benedetti, C. and Rossi, F. and Schroeder, C. B. and Esarey, E. and Leemans, W. P.},
  journal = {Phys. Rev. E},
  volume = {92},
  issue = {2},
  pages = {023109},
  numpages = {11},
  year = {2015},
  month = {Aug},
  publisher = {American Physical Society},
  doi = {10.1103/PhysRevE.92.023109},
  url = {https://link.aps.org/doi/10.1103/PhysRevE.92.023109}
}

\end{document}